\documentclass{aastex63}
\accepted{ApJ}
\shorttitle{Discovery of icy objects}
\shortauthors{Onaka et al.}

\begin{document}

\title{Discovery of two infrared objects with strong ice absorption in the  AKARI slit-less spectroscopic survey of the Galactic Plane}

\correspondingauthor{Takashi Onaka}
\email{onaka@astron.s.u-tokyo.ac.jp.jp}

\author[0000-0002-8234-6747]{Takashi Onaka}
\affiliation{Department of Physics, Faculty of Science and Engineering, 
Meisei University, 2-1-1 Hodokubo, Hino, Tokyo 191-8506, Japan\\
}
\affiliation{Department of Astronomy, Graduate School of Science, The University of Tokyo, 7-3-1 Hongo, Bunkyo-ku, Tokyo 113-0033, Japan\\
}

\author{Tomoyuki Kimura}
\altaffiliation{Present address: East Japan Railway Company, Tokyo 151-8578, Japan}
\affiliation{Department of Astronomy, Graduate School of Science, The University of Tokyo, 7-3-1 Hongo, Bunkyo-ku, Tokyo 113-0033, Japan\\
}

\author[0000-0001-7641-5497]{Itsuki Sakon}
\affiliation{Department of Astronomy, Graduate School of Science, The University of Tokyo, 7-3-1 Hongo, Bunkyo-ku, Tokyo 113-0033, Japan\\
}

\author[0000-0002-0095-3624]{Takashi Shimonishi}
\affiliation{Center for Transdisciplinary Research, Niigata University, Ikarashi-nihoncho 8050, Nishi-ku, Niigata 950-2181, Japan}
\affiliation{Environmental Science Program, Department of Science, Faculty of Science, Niigata University, Ikarashi-nihoncho 8050, Nishi-ku, Niigata 950-2181, Japan}



\begin{abstract}
We discover two infrared objects that show deep absorption features of H$_2$O, CO$_2$, and CO ices in the
{\it AKARI}/Infrared Camera (IRC) slit-less spectroscopic survey of the Galactic plane in 2.5--13\,$\mu$m.
Both objects are located neither in known star-forming regions nor in known dense clouds.  
For one of the objects, Object\,1, we successfully extract a spectrum from 2.5 to 13\,$\mu$m,
which also shows several absorption features in 5--13\,$\mu$m, including deep silicate absorption at 10\,$\mu$m.
For the other object, Object\,2, only a spectrum from 3.1 to 5\,$\mu$m is reliably extracted due to the presence of nearby overlapping objects
and faint nebulosity.
Both objects show warm ($>100$\,K) CO gas absorption in addition to the ice absorption features,
suggesting that they are embedded young stellar objects (YSOs).
On the other hand, both objects have spectral energy distributions (SEDs) that peak at around 5\,$\mu$m and decrease towards longer wavelengths.
These characteristics of the SEDs and the presence of deep absorption features cannot easily be accounted for by standard YSO models.
They may be explained as background stars behind dense clouds.  
We discuss possible nature of the objects and implications of the present discovery.

\end{abstract}



\section{Introduction} \label{sec:intro}
In cold, dense regions, various kinds of ice species are formed and play significant roles
in the interstellar chemistry as well as in the formation of stars and planetary systems \citep[for a review,][]{2015ARA&A..53..541B}.  
While {\it Infrared Space Observatory (ISO)} provided an extensive spectroscopic database of ice absorption features for massive young stellar
objects \citep[MYSO, e.g., ][]{2004ApJS..151...35G, 2005SSRv..119..293D} and CO$_2$ ice properties in the Taurus dark cloud
\citep[e.g,][]{1998ApJ...498L.159W, 2001ApJ...558..185N},
 {\it Spitzer} and {\it AKARI} as well as ground-based spectroscopy extended the study of ices to low-mass
young stellar objects (LYSO) and the various ice species in several dense clouds 
\citep[e.g.,][]{2004ApJS..154..359B, 2008ApJ...678..985B, 2011ApJ...729...92B,
2005ApJ...627L..33B, 2005ApJ...635L.145K, 2008ApJ...678.1005P, 2009ApJ...690..683R, 2011ApJ...740..109O, 2012A&A...538A..57A, 
2013ApJ...775...85N, 2017MNRAS.467.4753N}.  Theoretical and experimental studies suggest that
ices are formed on the grain surface via diffusive surface reactions or
energetic processes such as photolysis or radiolysis \citep[for recent reviews,][]{2013ChRv..113.8783H, 2017SSRv..212....1C,
2021SSRv..217...14M}.
Although the formation and evolution processes of ice species in dense regions are 
not yet fully understood, the profiles of absorption features in the infrared provide us
with valuable information on the thermal and energetic processes imposed upon ices as well as on the
nature of the objects associated with ice absorption
\citep[e.g.,][]{2003A&A...408..981P, 2011ApJ...740..109O, 2013ApJ...775...85N, 2015ARA&A..53..541B}.
The presence of ice absorption is also thought to be a reliable indicator for the identification of young
stellar objects \citep[YSOs,][]{2005MNRAS.364L..71V, 2008ApJ...686L..99S, 2010A&A...514A..12S, 2009ApJ...699..150S}, which is difficult to be made 
unambiguously solely by infrared photometric observations because some galaxies and dusty evolved stars have photometric characteristics similar to
YSOs
\citep[e.g.,][]{2008ApJ...674..336G, 2009ApJS..184..172G, 2010ApJ...720..679B, 
2012AJ....144..179K, 2014ApJ...791..131K}.

H$_2$O, CO$_2$, CO, and CH$_3$OH ices are known to be the major ice species in YSOs and dense clouds, and they have the major bands
at 3.0, 6.0, 13.1 (H$_2$O), 4.67 (CO), 4.26, 15.2\,$\mu$m  (CO$_2$), 3.53, 6.76,  8.9, and 9.75\,$\mu$m (CH$_3$OH) \citep[][and references therein]{1995A&A...296..810G,
2004ApJS..151...35G}. Other complex ice species also have characteristic bands in 4.6--10\,$\mu$m \citep
{2004ApJS..151...35G, 2015ARA&A..53..541B}.  Therefore, the near-infrared (NIR) and mid-infrared (MIR) spectroscopy is an efficient
means to study the properties of interstellar ices as well as the nature of objects associated with ice absorption.

In this paper, we report discovery of two interesting infrared objects that show strong
absorption bands of H$_2$O ice at 3.0\,$\mu$m, CO$_2$ ice at 4.26\,$\mu$m, and CO ice at 4.67\,$\mu$m based on a NIR to MIR
slit-less spectroscopic survey (2.5--13\,$\mu$m) of the Galactic plane carried out with the Infrared Camera (IRC) onboard the {\it AKARI} satellite
\citep{2007PASJ...59S.401O}.  Both objects are located neither in known star-forming
regions nor in known dense clouds.  They could be run-away YSOs \citep{2005A&A...437..247D,
2019A&A...624A..66R}, which elude past YSO surveys,  or located behind unknown compact, dense clouds.

In \S\ref{sec:obs}, the observations and data reduction are described, and the results are
presented in \S\ref{sec:results}.  Analysis of the spectra is given in \S\ref{sec:analysis}.
Possible identification of the nature of the objects is discussed in \S\ref{sec:discussion} and
a summary is given in \S\ref{sec:summary}.

\section{Observations and data reduction} \label{sec:obs}

The present observations were carried out in the slit-less spectroscopic survey mode with the grisms of the IRC,
which had a field-of-view of $10\arcmin \times 10\arcmin$ and obtained spectra of point sources for 2.5--13\,$\mu$m
\citep{2007PASJ...59S.411O},
as part of the program of the Interstellar Medium in our Galaxy and Nearby galaxies 
\citep[ISMGN,][]{2009AdSpR..44.1038K}.   Spectra of 2.5--5\,$\mu$m were taken with the NIR channel of
the IRC in the NIR Grism (NG) mode, while those of 5--13\,$\mu$m were obtained with the MIR-S channel in the 
Short-MIR Grism 1 (SG1, 5.0--8.2\,$\mu$m)
and Short-MIR Grism 2 (SG2, 7.6--13.0\,$\mu$m) modes.  The spectral resolutions are about 0.03, 0.12, and 0.21\,$\mu$m for the NG,
SG1, and SG2 modes, respectively \citep{2007PASJ...59S.411O}.  We surveyed nine regions in the Carina arm, eight
in the Crux arm, and five in the Perseus arm.  Table~\ref{tab:target} summarizes the positions, IDs, and dates of the observations.

\begin{deluxetable*}{lclrl}
\tablecaption{Observation log of the slit-less spectroscopic survey of the Galactic plane\label{tab:target}}
\tablewidth{0pt}
\tablehead{
\colhead{Target name} &  \colhead{Observation ID} & \multicolumn{2}{c}{Position\tablenotemark{a}} & \colhead{Observation date}\\
 & &  \multicolumn{2}{c}{R.A.(J2000.0) Dec.}
}
\startdata
CAR-079\_O001& 1402304.1 & 09$^{\mathrm{h}}$53$^{\mathrm{m}}$29$^{\mathrm{s}}$ &  -56$\degr$14$\arcmin$00$\arcsec$   &  2007 June 29\\
CAR-079TO002& 1400201.1 & 09$^{\mathrm{h}}$55$^{\mathrm{m}}$42$^{\mathrm{s}}$ &  -55$\degr$50$\arcmin$29$\arcsec$ & 2006 December 29 \\
CAR-079TS004& 1400203.1 & 09$^{\mathrm{h}}$57$^{\mathrm{m}}$53$^{\mathrm{s}}$ &  -55$\degr$26$\arcmin$49$\arcsec$ &  2006 December 29  \\
CAR-079TO003& 1400197.1 & 10$^{\mathrm{h}}$01$^{\mathrm{m}}$24$^{\mathrm{s}}$&  -56$\degr$27$\arcmin$04$\arcsec$  &  2006 December 31\\
CAR-079\_S002& 1402301.1 & 10$^{\mathrm{h}}$04$^{\mathrm{m}}$19$^{\mathrm{s}}$ &  -56$\degr$44$\arcmin$58$\arcsec$  &  2007 July 02\\
CAR-057TO005& 1400229.1 & 13$^{\mathrm{h}}$13$^{\mathrm{m}}$55$^{\mathrm{s}}$ &  -62$\degr$45$\arcmin$34$\arcsec$  &  2007 February 06 \\
CAR-057TO003& 1400225.1 & 13$^{\mathrm{h}}$14$^{\mathrm{m}}$18$^{\mathrm{s}}$ &  -63$\degr$15$\arcmin$27$\arcsec$  &  2007 February 07\\
CAR-057TT001& 1400231.1 & 13$^{\mathrm{h}}$20$^{\mathrm{m}}$44$^{\mathrm{s}}$&  -65$\degr$11$\arcmin$49$\arcsec$ &  2007 February 06\\
CAR-057TS002& 1400215.1 &13$^{\mathrm{h}}$22$^{\mathrm{m}}$35$^{\mathrm{s}}$ & -62$\degr$39$\arcmin$19$\arcsec$ &  2007 February 10 \\
CRU-048TO003 & 1400259.1 & 14$^{\mathrm{h}}$03$^{\mathrm{m}}$25$^{\mathrm{s}}$ &  -61$\degr$10$\arcmin$09$\arcsec$  & 2007 February 12\\
CRU-048TO002 & 1400257.1 & 14$^{\mathrm{h}}$04$^{\mathrm{m}}$34$^{\mathrm{s}}$ &  -61$\degr$38$\arcmin$59$\arcsec$ & 2007 February 13  \\
CRU-048TO004 & 1400261.1 & 14$^{\mathrm{h}}$07$^{\mathrm{m}}$23$^{\mathrm{s}}$ &  -61$\degr$01$\arcmin$38$\arcsec$ & 2007 February 13 \\
CRU-048TS001 & 1400247.1 & 14$^{\mathrm{h}}$09$^{\mathrm{m}}$57$^{\mathrm{s}}$&  -61$\degr$59$\arcmin$24$\arcsec$ & 2007 February 14 \\
CRU-048TS003 & 1400251.1 & 14$^{\mathrm{h}}$13$^{\mathrm{m}}$53$^{\mathrm{s}}$ &  -61$\degr$49$\arcmin$49$\arcsec$  & 2007 February 15 \\
CRU+032\_SP05 & 1401027.1 & 18$^{\mathrm{h}}$48$^{\mathrm{m}}$41$^{\mathrm{s}}$ &  -00$\degr$14$\arcmin$45$\arcsec$  & 2007 April 02\\
CRU+032\_S003 & 1401033.1 & 18$^{\mathrm{h}}$59$^{\mathrm{m}}$30$^{\mathrm{s}}$ &  -01$\degr$44$\arcmin$43$\arcsec$  & 2007 April 06\\
CRU+032\_SP01 & 1401025.1 & 18$^{\mathrm{h}}$59$^{\mathrm{m}}$32$^{\mathrm{s}}$ &  -01$\degr$36$\arcmin$49$\arcsec$  & 2007 April 06\\
PER+070\_S001 & 1401019.1 & 20$^{\mathrm{h}}$04$^{\mathrm{m}}$51$^{\mathrm{s}}$ &  29$\degr$12$\arcmin$08$\arcsec$  & 2007 May 02\\
PER+070\_O001 & 1401017.1 & 20$^{\mathrm{h}}$10$^{\mathrm{m}}$45$^{\mathrm{s}}$ &  32$\degr$35$\arcmin$26$\arcsec$  & 2007 May 06\\
PER+070\_S002 & 1402365.1 & 20$^{\mathrm{h}}$12$^{\mathrm{m}}$59$^{\mathrm{s}}$ &  33$\degr$30$\arcmin$10$\arcsec$  & 2007 May 08\\
PER+070\_O003 & 1402370.1 & 20$^{\mathrm{h}}$14$^{\mathrm{m}}$02$^{\mathrm{s}}$ &  32$\degr$43$\arcmin$58$\arcsec$ &   2007 May 08\\
PER+070\_O002 & 1402369.1 & 20$^{\mathrm{h}}$15$^{\mathrm{m}}$21$^{\mathrm{s}}$ &  33$\degr$08$\arcmin$57$\arcsec$  & 2007 May 08\\
\enddata
\tablenotetext{a}{The intended center position of the field-of-view}
\end{deluxetable*}

\begin{deluxetable*}{ccc}
\tablecaption{2MASS ID, position, and photometric data of the objects\tablenotemark{a}  \label{tab:location}}
\tablewidth{0pt}
\tablehead{
\colhead{Properties/Band} & \colhead{Object\,1} & \colhead{Object\,2}}
\startdata
2MASS ID &J14041323-6112401 & J14042016-6115495 \\
R.A. (J2000.0) DEC & 14$^{\mathrm{h}}$04$^{\mathrm{m}}$13.2$^{\mathrm{s}}$ \,  -61$\degr$12$\arcmin$40$\farcs$1  & 14$^{\mathrm{h}}$04$^{\mathrm{m}}$20.2$^{\mathrm{s}}$ \,  -61$\degr$15$\arcmin$49$\farcs$5  \\
$J$ & $> 16.712$ & $> 17.548 $\\
$H$ & $14.914 \pm 0.11$ & $> 16.880$  \\
$K_\mathrm{s}$ & $10.346 \pm 0.023$ & $11.730 \pm 0.026$\\
$W1$ & $8.755 \pm 0.023$ & $9.600 \pm 0.024$ \\
$W2$ & $6.828 \pm 0.020$ & $7.903 \pm 0.020$ \\
$W3$ & $6.699 \pm 0.034$ & $7.216 \pm 0.040$ \\
$W4$ & $5.772 \pm 0.066$ & $> 4.615$\tablenotemark{b} \\
$[3.6]$ & $7.676 \pm 0.041$ & $8.493 \pm 0.040$ \\
$[4.5]$ & $6.592 \pm 0.056 $& $7.888 \pm 0.037$ \\
$[5.8]$ & $6.050 \pm 0.036$ & $7.151 \pm 0.037$ \\
$[8]$ & $5.925 \pm 0.030$ & $7.170 \pm 0.029$ \\
$[24]$ & $4.65 \pm 0.02$ & $6.00 \pm 0.33$ \\
{\it AKARI} 9\,$\mu$m (mJy) & $119.3 \pm 14.6$ & $< 50$ \\
\enddata
\tablenotetext{a}{Photometric data are given in magnitude except for {\it AKARI} 9\,$\mu$m, which is in units of mJy.  
The {\it WISE} data are taken from the ALLWISE catalog and the IRAC and MIPS data are from the GLIMPSE and  MIPSGAL catalogs.
The {\it AKARI} data are taken
from the JAXA/ISAS\ server (see text).} 
\tablenotetext{b}{We set this value as a lower limit because of the presence of faint nebulosity (see \S\ref{sec:results}).} 
\end{deluxetable*}

\begin{figure}[ht!]
\epsscale{0.7}
\plotone{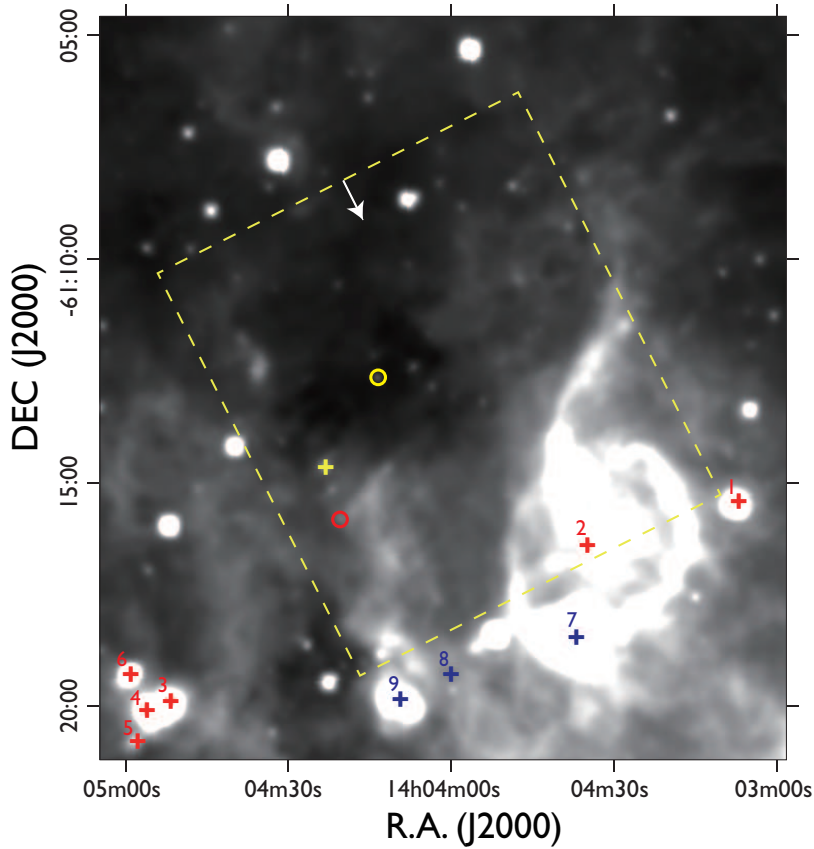}
\caption{Locations of the two objects that show strong ice absorption features in the {\it AKARI} slit-less survey.  
The background is the WISE band 3 (12\,$\mu$m) image. 
The yellow dashed rectangle shows the field-of-view of the IRC observations ($10\arcmin \times 10\arcmin$), while the white arrow indicates the upward dispersion direction
in the slit-less spectrum images shown in Figure~\ref{fig:slit}.   The yellow and red circles show the locations of  Object\,1 and Object\,2, respectively.  
The yellow plus indicates an AGB star candidate \citep[2MASS\, J14042291-6114400,][]{2008AJ....136.2413R}, the red pluses show \ion{H}{2} regions \citep[1: G311.440+00.424, 2:
G311.489+00.394, 3:  G311.621+00.295, 4: G311.629+00.289, 5: G311.629+00.277, 6; G311.638+00.30,][]{2017A&A...597A.114B}, and the blue pluses display
star-forming regions \citep[7: IRAS\,14000-6104, 8: IRAS\,14004-6104, 9: RAGL\,4188,][]{2002ARep...46..193A}.
 \label{fig:location}}
\end{figure}

\begin{figure}[ht!]
\plotone{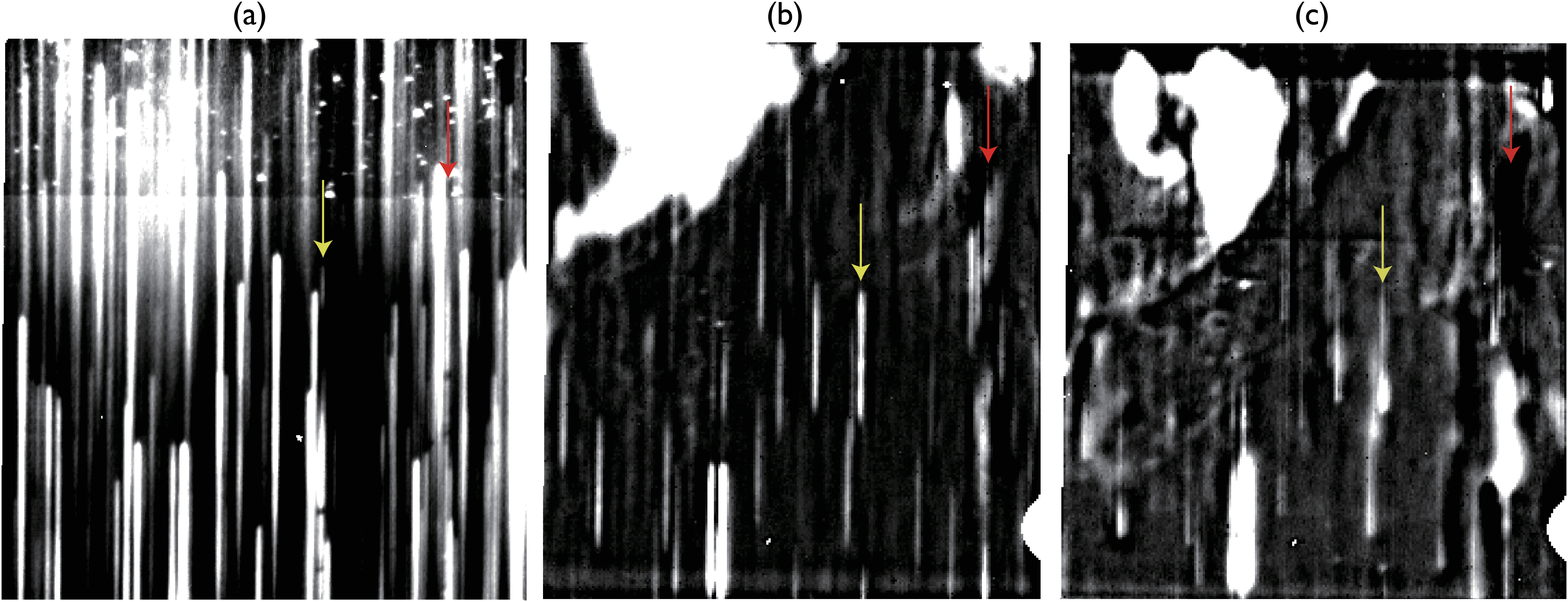}
\caption{Slit-less images of the present observations.  (a): NG image (2.5--5\,$\mu$m), (b): MIR-SG1 image (5.0--8.2\,$\mu$m), and (c): MIR-SG2 image (7.6--13.0\,$\mu$m).  
The yellow and red arrows indicate Object\,1 and Object\,2, respectively. 
The wavelength decreases upward in the NG image, while increases in the SG1 and SG2 images. 
The two black stripes seen in the NG spectra of the two objects in (a) correspond to the absorption bands of CO$_2$ ice at 4.26\,$\mu$m and CO ice at 4.67\,$\mu$m.  For Object\,1,
the broad absorption of H$_2$O ice at 3.0\,$\mu$m is also visible, while that of Object\,2 is not clearly seen due to the nearby bright source.} \label{fig:slit}
\end{figure}

Two objects are found in one of the Crux arm
regions (target name of CRU-048TO003), which  show deep absorption of CO$_2$ ice at 4.26\,$\mu$m and CO ice at 4.67\,$\mu$m in their slit-less spectra of the NIR channel by visual inspection.  
We call the two objects Object\,1 and Object\,2 in the following. 
No similar objects that show strong ice absorption were found in the other 22 regions by visual inspection.
Their locations are indicated in Figure~\ref{fig:location} on the WISE band 3 (12\,$\mu$m) image together with
nearby \ion{H}{2} regions (red pluses), star-forming regions (blue pluses), and an Asymptotic Giant Branch (AGB) star candidate (yellow plus).  Figure~\ref{fig:location} shows that Object\,1 is located
\replaced{
in a moderately dark region in the MIR and}{in a region that is associated neither with bright infrared sources nor with strong diffuse emission in the MIR.  
No appreciable dark clouds are seen in the optical extinction map
\citep{2005PASJ...57S...1D}, while}
Object\,2 is located in a region associated with faint diffuse emission.  Nearby star-forming activities are located at $> 4\arcmin$ south of the two objects.
The slit-less spectrum images are shown in Figure~\ref{fig:slit}.  The NG image (Figure~\ref{fig:slit}a) shows the
presence of two black stripes that correspond to the CO$_2$ ice and CO ice absorption bands in both objects.  For Object\,1, the broad H$_2$O ice absorption at 3.0\,$\mu$m is also recognized.
Both objects are listed in the point source catalogs of the Two Micron All Sky Survey (2MASS) \citep{2006AJ....131.1163S}, 
the Wide-field Infrared Survey Explorer (WISE) \citep{2010AJ....140.1868W}, the {\it Spitzer} 
Galactic Legacy Infrared Mid-Plane Survey Extraordinaire (GLIMPSE) survey \citep{2003PASP..115..953B, 2009PASP..121..213C}, 
and the {\it Spitzer} MIPS Galactic (MIPSGAL) survey \citep{2009PASP..121...76C}.  Object\,1 is also listed in the point source catalog of {\it AKARI}
MIR all-sky survey \citep{2010A&A...514A...1I}, while Object\,2 is not.  We confirmed that Object\,2 had been observed in the  {\it AKARI}
MIR all-sky survey and we gave the nominal upper limit of 50\,mJy for it at 9\,$\mu$m \citep[][see also \S\ref{sec:results}]{2010A&A...514A...1I}.  
In the following, we denote  {\it WISE} band 1--4 data as $W1$--$W4$,
IRAC band 1--4 data as [3.6], [4.5], [5.8], and [8], and MIPS\,24\,$\mu$m data as [24] in magnitude.  The positions, 2MASS IDs, and photometric data of the two objects are summarized in
Table~\ref{tab:location},  where
the {\it WISE} data are taken from the ALLWISE catalog and the IRAC and MIPS data are from the GLIMPSE and  MIPSGAL catalogs
in the NASA/IPAC Infrared Science Archive (IRSA)\footnote{ \url{https://irsa.ipac.caltech.edu/frontpage/}}.  The {\it AKARI} data are taken
from the JAXA/ISAS\ server\footnote{ \url{https://www.ir.isas.jaxa.jp/AKARI/Archive/Catalogues/PSC/}}.
The photometric data are given in magnitude except for {\it AKARI} 9\,$\mu$m, which is in units of mJy.


Both objects are located near the Galactic plane ($b \sim 0\fdg4$).
Because of the presence of nearby sources, they do not satisfy the criteria
 of the automatic selection and are not included in the {\it AKARI} MIR slit-less spectroscopic catalog \citep{2019PASJ...71....3Y}.
We applied a careful data reduction procedure to avoid contamination from nearby sources
 and extracted the spectra of both objects from the IRC
slit-less spectral images of the NIR and MIR-S channels by using the IRC spectroscopic data reduction toolkit of version 
20150331\footnote{ \url{https:///www.ir.isas.jaxa..jp/AKARI/Observation/support/IRC\#software}} and
the latest wavelength calibration for the NIR spectrum was applied \citep{2016PASJ...68...27B}.  

In the catalog production process, MIR
spectra are extracted for 7 pixels in the spatial direction to not lose the source flux and obtain an optimal signal-to-noise ratio \citep{2019PASJ...71....3Y}.
We needed to apply a narrower spatial extraction window to avoid contamination in the spectrum extraction of the present data and 
the aperture corrections were applied based on the point spread function for the narrow extraction window
\citep{2007PASJ...59S.411O}.  The sky background was subtracted using the sky data near the object.
Object\,1 is located in a relatively empty region of the sky on the Galactic plane, and there is an infrared source located at a distance of about $8\arcsec$ from Object\,1 
in the direction vertical to the dispersion (Figure~\ref{fig:location}).  
The image quality (FWHM) of the NIR channel is better than $4\farcs3$ \citep{2007PASJ...59S.401O}.  The NG spectrum (2.5--5\,$\mu$m) of Object\,1 was extracted for
4 pixels ($= 5\farcs84$) to avoid the contamination from the nearby source.  We, then, applied a two-Gaussian component fit in the spatial direction developed by \citet{2012SPIE.8442E..44S} to
estimate the contamination of nearby sources \citep[see also][]{2013ApJ...775...85N}.
We confirmed that the spectrum extracted by the toolkit agreed with the two-Gaussian component fit within the uncertainty and the contamination was insignificant.
We extracted the SG1 and SG2 spectra (5--13\,$\mu$m) for 3 pixels ($= 7\farcs02$).   The image quality (FWHM) of the MIR-S channel is better than $5\farcs1$ \citep{2007PASJ...59S.401O}.
The nearby source is fainter in the MIR (Figure~\ref{fig:slit}) and does not make a contamination to the SG1 and SG2 spectra of Object\,1.
The spectra of SG1 and SG2 are smoothly connected at 8\,$\mu$m and
we use the SG1 spectrum for wavelengths shorter than 8\,$\mu$m and the SG2 spectrum for those longer than 8\,$\mu$m.

Around Object\,2, there are a nearby source with a similar brightness at
a distance of $5\arcsec$ and a source brighter than Object\,2 at a distance of $8\arcsec$ in the direction vertical to the dispersion in the NG image (Figure~\ref{fig:slit}a).
We applied a two-Gaussian component fit in the spatial direction to estimate the contamination from the nearby sources in the NG spectrum and adopted 
a narrow extraction window of 2 pixels ($=2\farcs92$), 
which reduced the signal-to-noise ratio but enabled to avoid the contamination for
the spectral region 3.1--5.0\,$\mu$m.  
We found that it was not possible to reliably recover the spectrum for 2.5--3.1\,$\mu$m due to the faint continuum with
the strong H$_2$O ice absorption at 3\,$\mu$m.
Object\,2 is also located in a region with faint nebulosity in the MIR (Figure~\ref{fig:location}) and is very faint in the SG2 image.
In the SG1 image, it was impossible to avoid contamination from the nebulosity
completely and obtain a reliable spectrum even with a narrow extraction window.  
It was also difficult to select a reliable sky area for the background subtraction.
To recover spectral information as much as possible, therefore, 
the same window size of 3 pixels as for Object\,1 was adopted.  The spectrum for 5--8\,$\mu$m taken with the SG1 mode was able to be extracted
with low reliability, while the spectrum taken with the SG2 mode (8--13\,$\mu$m) was very faint and could not be extracted.
Therefore, Object\,2 has a reliable spectrum extracted only for 3.1--5.0\,$\mu$m.

During the slit-less observations, a 3\,$\mu$m image and a 9\,$\mu$m image were taken to provide the origin of the wavelength scale in the slit-less images
\citep{2007PASJ...59S.411O}.  The absolute wavelength in the spectrum extraction process 
is estimated to be accurate better than a half pixel, which corresponds to 0.005, 0.03, and 0.05\,$\mu$m for
the NG, SG1, and SG2 spectra, respectively.
Lastly, the spectrum between 4.95--5.0\,$\mu$m in the NG was removed because of the
potential contamination of the second-order light  \citep{2016PASJ...68...27B}.

\section{Results} \label{sec:results}

\begin{figure}[ht!]
\plotone{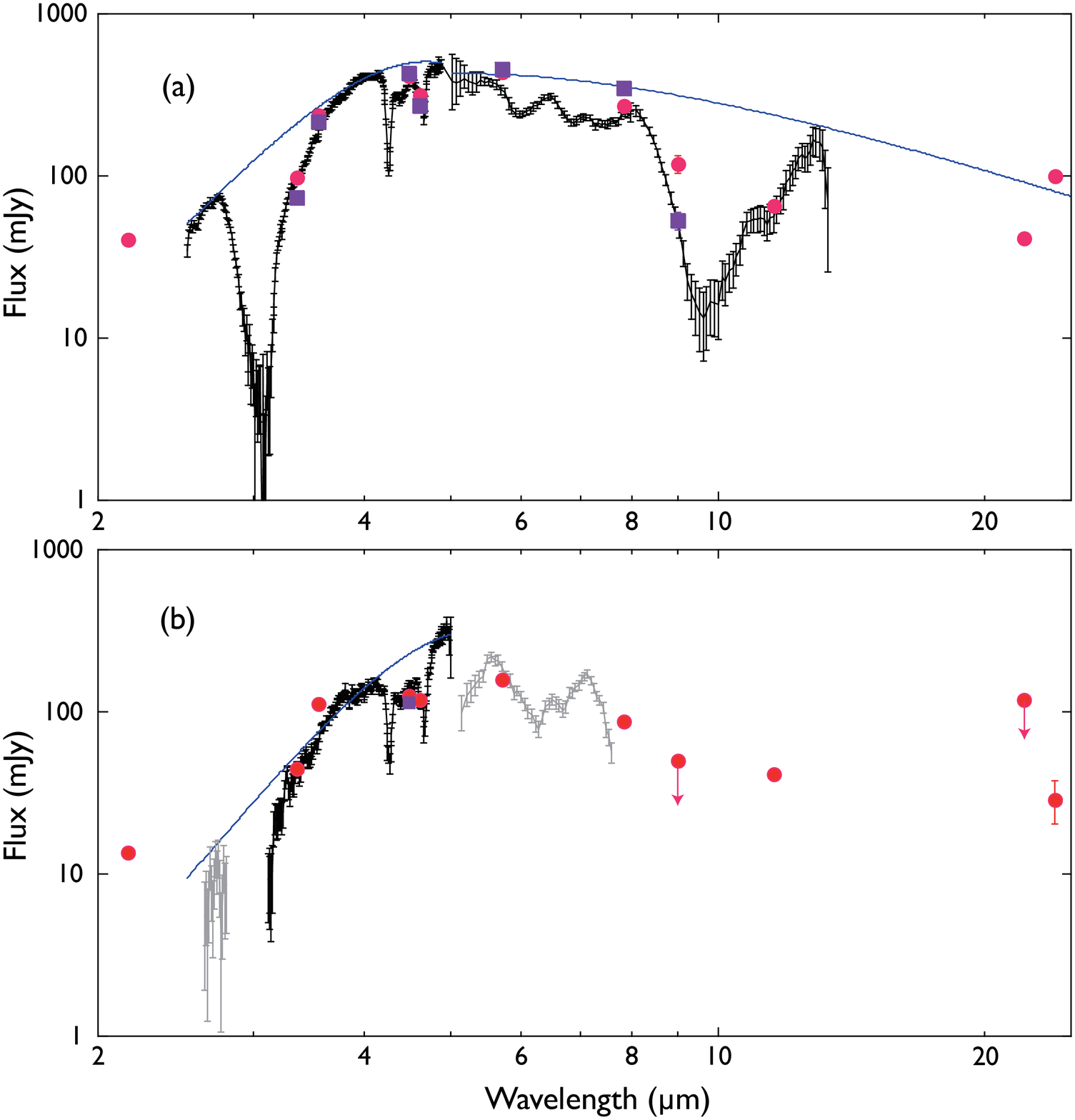}
\caption{Observed {\it AKARI}/IRC spectra of Object\,1 (a) and Object\,2 (b). The red circles show the photometric data points
from the catalogues,
while the purple squares indicate the color-corrected ones, for which the spectral information is derived from
the observed spectra.  The error bars are also shown for the photometric data, but are smaller than the size of symbols for most points.
The blue solid lines indicate the assumed continua (see \S\ref{sec:analysis}). In (b), the grey lines indicate the spectral
regions, which are not reliable due to the possible contamination from nearby sources and faint nebulosity. \label{fig:spectra}}
\end{figure}

The extracted spectra of Object\,1 and Object\,2 are shown in Figure~\ref{fig:spectra}.
For Object\,2, the gray lines indicate the regions, where the spectra are affected by nearby
sources and faint nebulosity, and are not reliable.  There is a small jump between the NG and SG1 spectra of Object\,1
at 5\,$\mu$m.  We did not make any correction to stitch the spectra.
Photometric data of 2MASS, {\it Spitzer}/IRAC and MIPS, {\it WISE}, and {\it AKARI} are also plotted together. 
The red circles show the original catalog values,
while the purple squares indicate the color-corrected ones, for which the spectral information
for color correction is estimated from the observed spectrum.  For Object\,1 (Figure~\ref{fig:spectra}a), 
the color corrections are generally small except for
those at the deep absorption features ($W1$ (3.4\,$\mu$m), IRAC band 4 (8\,$\mu$m), and 
{\it AKARI} 9\,$\mu$m).  For Object\,2 (Figure~\ref{fig:spectra}b), 
the color correction is possible only for IRAC band 2 (4.5\,$\mu$m).   The photometric data are generally in agreement
with the observed spectra, suggesting that the spectra are extracted properly.

For Object\,1, deep absorption features of H$_2$O ice at 3\,$\mu$m and silicate at
10\,$\mu$m are apparent, which can be noticed even in the photometric data.  The absorption features of CO$_2$ ice at 4.26\,$\mu$m
and CO ice at 4.67\,$\mu$m are clearly seen.  In addition, several absorption bands are present in 5--8\,$\mu$m.
They are discussed in the next section.  In the 20\,$\mu$m region, the $W4$ (22\,$\mu$m)
and MIPS 24\,$\mu$m show discrepancy.  Its cause is not clear at the present.  Both data, however, suggest that the
flux is decreasing towards longer wavelengths.

For Object\,2, the spectrum at around 3\,$\mu$m is very faint and is not reliably extracted as described in \S\ref{sec:obs}.  However,
the presence of deep absorption is suggested by the faintness of the flux at around 3\,$\mu$m
and  by the decreasing trend of the spectrum from $\sim 3.3$\,$\mu$m.  The spectrum of 3.1--5.0\,$\mu$m shows
several absorption features, including CO$_2$ ice at 4.26\,$\mu$m and CO ice at 4.67\,$\mu$m.
The $W4$ (22\,$\mu$m) data are appreciably brighter than the MIPS 24\,$\mu$m
data.  Object\,2 is associated with faint nebulosity in the MIR region (Figure~\ref{fig:location}).  We attribute the difference
to the larger beam of {\it WISE}, which is affected by the nebulosity, and take the {\it WISE} data point as an upper limit.  
The nebulosity also affects detection of {\it AKARI} 9\,$\mu$m.  Inspection of the original data confirms 50\,mJy as 
a conservative upper limit \citep{2010A&A...514A...1I}, which may suggest the presence of silicate absorption.
The spectrum of Object\,2 at wavelengths longer than
20\,$\mu$m is not well constrained, and only a decreasing trend similar to Object\,1 is suggested.

In addition to the NIR and MIR photometry, we searched for far-infrared (FIR) data of both objects
in the {\it Herschel} Infrared Galactic Plane Survey data \citep[HIGAL,][]{2016A&A...591A.149M}.
There are no objects listed at the object positions in the HIGAL catalogs.\footnote{ \url{https://tools.ssdc.asi.it/HiGAL.jsp}}
Upper limits of the detection depend on the position of the sky.  From the fluxes of objects detected within $2\arcmin$ of the
target positions and
the completeness limits of the survey \citep{2016A&A...591A.149M},
we conservatively estimate upper limits as 0.8, 3, 9, 10, and 10\,Jy at 70, 160, 250, 350, and 500\,$\mu$m,
respectively.

\section{Absorption features} \label{sec:analysis}

Because of the small jump between NIR and MIR spectra, we define the
continuum separately for the NIR and MIR regions.  For 2.5--5\,$\mu$m, 
we fit a quadratic polynomial to the logarithmic of the flux at the continuum regions, 
2.58--2.65, 3.86--4.15, and 4.73--4.82\,$\mu$m for Object\,1, and 3.51--4.13 and 4.78--4.93\,$\mu$m for
Object\,2.  These continua fit the continuum regions adequately.  
For wavelengths longer than 5\,$\mu$m, there are no reliable continuum points and we simply
fit a 1000K blackbody for Object\,1 \citep{2004ApJS..151...35G} to the photometry points of IRAC bands 3 and 4,
and MIPS 24\,$\mu$m.
No continuum is estimated for Object\,2 for wavelengths longer than 5\,$\mu$m.
The assumed continua are shown by the blue solid lines in Figure~\ref{fig:spectra}.  The optical depth is estimated
using these continua.  In the following, we discuss the spectral features in 2.5--4.0, 4.0--5.0, 5.0--8.0, and
8--13\,$\mu$m, separately.  For Object\,2, only the features in the reliable spectrum of 4.0--5.0\,$\mu$m are discussed.

\subsection{2.5--4.0\,$\mu$m}\label{subsec:2-4}
\begin{figure}[ht!]
\plotone{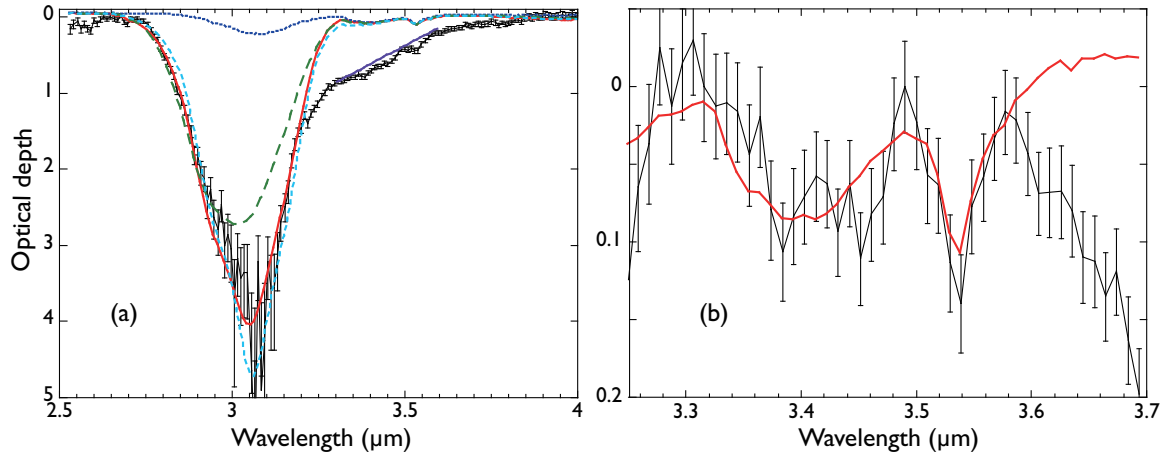}
\caption{ Optical depth spectrumc of Object\,1 in the 2.5--4.0\,$\mu$m region.  The observed spectrum is shown by the black solid lines with
the error bars.  (a) The solid red line indicates the fit with the amorphous H$_2$O ice at 60\,K, while the green long-dashed line and the light-blue short-dashed line 
show the fits with the amorphous H$_2$O ices at 15\,K and 100\,K, respectively \citep{2009ApJ...701.1347M}. The blue dotted line indicates the contribution of CH$_3$OH ice
estimated from the feature at 3.54\,$\mu$m (see Figure b and text).  The solid purple line shows the assumed linear baseline for the extraction of the
CH$_3$OH feature shown in b. (b) A close-up view of the 3.25--3.70\,$\mu$m region after subtraction of
the baseline.  The red solid line indicates the
fit with the spectrum of CH$_3$OH ice at 15K \citep{2004AdSpR..33...14F}.  \label{fig:Obj1:2-4}}
\end{figure}

Figure~\ref{fig:Obj1:2-4} shows the spectrum of Object\,1 in the 2.5--4\,$\mu$m region.  In addition to the
deep absorption of H$_2$O ice at 3.0\,$\mu$m, there are weak features seen at $\sim 3.4$ and 3.54\,$\mu$m.
They can be attributed to O-H stretching modes of CH$_3$OH ice \citep{1986A&AS...64..453D}.   CH$_3$OH ice
also has a stronger band at 3.07\,$\mu$m.
We first fit the features in 3.3--3.6\,$\mu$m with CH$_3$OH ice and estimate its contribution to the
3.0\,$\mu$m band.  
A linear baseline (solid purple line in Figure~\ref{fig:Obj1:2-4}a) is subtracted from the observed spectrum and the spectrum in 3.3--3.6\,$\mu$m 
is fitted with the absorbance data of pure CH$_3$OH ice at 15\,K \citep{2004AdSpR..33...14F}
taken from the Leiden laboratory ice database.\footnote{\url{https://icedb.strw.leidenuniv.nl}}
The band profile hardly changes below 100\,K and the choice of different temperature data does not affect the fit result.
Pure CH$_3$OH ice data fit the observed spectrum reasonably well (Figure~\ref{fig:Obj1:2-4}b).  From the fit, we estimate
the column density of CH$_3$OH ice as $(2.6 \pm 0.6) \times 10^{17}$\,cm$^{-2}$, assuming that the band strength of the
3.54\,$\mu$m  band is 
$7.6 \times 10^{-18}$\,cm\,molecule$^{-1}$ \citep{1986A&AS...64..453D}.  With the estimated column density, the contribution
of  CH$_3$OH ice to the 3\,$\mu$m absorption is found to be insignificant (blue dotted line in Figure~\ref{fig:Obj1:2-4}a).

Figure~\ref{fig:Obj1:2-4}b also suggests the presence of weak features at around 3.4 and 3.47\,$\mu$m.  
The 3.47\,$\mu$m feature has been
seen in MYSOs, LYSOs, and background stars \citep[][reference therein]{2016A&A...585A.107S}.  It has been attributed to
either a C\sbond H vibration mode of hydrogen atoms bonded to tertiary carbon atoms \citep{1992ApJ...399..134A} or to ammonia hydrate formed in 
NH$_3$:H$_2$O mixture ice \citep{2001A&A...365..144D, 2002A&A...394.1057D}.  
The 3.4\,$\mu$m feature has often been seen in diffuse clouds and could be a sign of hydrocarbons \citep{1991ApJ...371..607S, 1994ApJ...437..683P,
2013ApJ...770...78C}. \added{The laboratory data of CH$_3$OH fit the 3.4\,$\mu$m feature fairly well, and there is no need for a 3.4\,$\mu$m hydrocarbon
feature.  It should, however, be noted that the fit is also dependent on
the assumed continuum.}
Further observations with a higher signal-to-noise ratio
are needed to confirm their presence.

A strong extended red wing on the 3.0\,$\mu$m H$_2$O ice absorption band is clearly seen in Figure~\ref{fig:Obj1:2-4}a.  While a correlation of the red wing with the
column density of H$_2$O ice is suggested for YSOs \citep{2006A&A...449..251T}, there are objects that show the 3.0\,$\mu$m band without the red wing and those with
a very strong wing \citep{2013ApJ...775...85N}.  The origins of the red wing are still in debate.  
The red wing and its references are discussed in more detail in \citet{2013ApJ...775...85N} and \citet{2015ARA&A..53..541B}.  
The profile of the red wing of Object\,1 is similar to Type 1 in \citet{2013ApJ...775...85N}, being typical of
YSOs and background stars.  
In this paper, we do not attempt to fit the red wing.  We also avoid the region around the peak of the 3.0\,$\mu$m absorption in the fit because
of the low signal-to-noise ratio.  We fit the spectrum of the regions of 2.7--2.9\,$\mu$m and 3.1--3.2\,$\mu$m with the spectrum of amorphous H$_2$O ice.  
We adopt the optical constants of amorphous H$_2$O ice measured at various temperatures between 20 and 150\,K provided  by \citet{2009ApJ...701.1347M}.
Because the 3.0\,$\mu$m band is strong and broad, it is necessary to take account of the particle shape effect \citep{1997A&A...328..649E, 2013ApJ...775...85N}.
We assume a continuous distribution of ellipsoids (CDE) for the shape.  The original CDE assumes a flat distribution
of the shape, giving equal probabilities to extreme shapes, i.e., infinitely narrow needles or thin disks \citep{1983uaz..rept.....B}.  In this paper, we assume a shape
distribution with a peak at a sphere \citep{1992A&A...261..567O} as a more realistic distribution.  The difference in the assumed distribution is
small and does not affect the following results.

The amorphous H$_2$O ice at 60\,K is found to show the best fit to the observed spectrum (red solid line in Figure~\ref{fig:Obj1:2-4}a).  The
amorphous H$_2$O ice at 15\,K has a peak at a shorter wavelength and cannot fit the longer-wavelength side of the observed profile very well (green long-dashed line), while
the one at 100\,K shows a sharper peak due to partial crystallization, which does not fit the observation either (light-blue short-dashed line).  
Note that the fits have a larger weight on the 2.7--2.9\,$\mu$m because of the higher signal-to-noise ratio than in 3.1--3.2\,$\mu$m.
The present analysis suggests that the H$_2$O ice
towards Object\,1 is accounted for by amorphous ice, which is thermally processed to some extent.
The column density of H$_2$O ice is estimated from the integrated band strength as $(54.9 \pm 1.1) \times 10^{17}$\,cm$^{-2}$,
 assuming that the band strength is given as $2.1 \times 10^{-16}$\,cm\,molecule$^{-1}$ for the
 amorphous H$_2$O ice at 60\,K \citep{2009ApJ...701.1347M}.
	
For Object\,2, the spectrum between 3.1 and 4.0\,$\mu$m is extracted without contamination.  However, this spectral range of Object\,2 is much fainter than that of Object\,1 
(Figure~\ref{fig:spectra}) and thus noisy.  No features are clearly seen in this spectral range and we do not discuss it in this paper.

\subsection{4.0--5.0\,$\mu$m}\label{subsec:45}

\begin{figure}[ht!]
\plotone{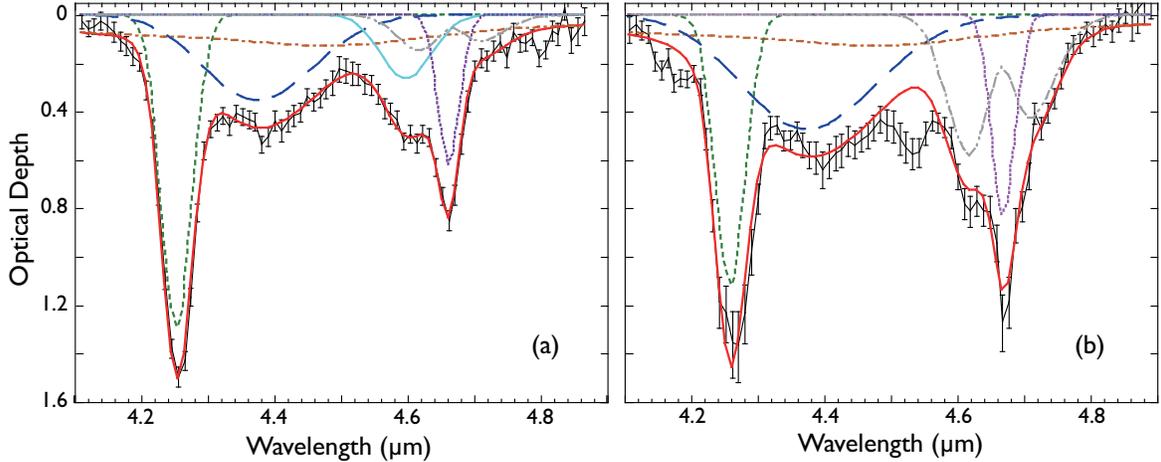}
\caption{ Optical depth spectra at 4.0--5.0\,$\mu$m  for Object\,1 (a) and Object\,2 (b). The observed spectra are indicated by the
black solid lines with the error bars and the best fit results are shown by the red solid lines.  The green short-dashed lines and the purple dotted
lines indicate the CO$_2$ and CO ice components, respectively.  
The gray double-dotted dashed lines show the CO gas component and the blue long dashed lines show the unidentified
4.4\,$\mu$m component.  The light blue solid line indicates the XCN component for
Object\,1, which is not included for Object\,2 (see text).
The brown dot-dashed lines show the contribution from H$_2$O ice, which is  estimated from the
3.0\,$\mu$m absorption for Object\,1.  For Object\,2, we assume the same contribution as in Object\,1 (see text).} \label{fig:CO2}
\end{figure}

Figure~\ref{fig:CO2} shows the 4.0--5.0\,$\mu$m spectra of the optical depth for Object\,1 and Object\,2, which indicate the presence of
several absorption features.  H$_2$O ice is known to have a broad, shallow feature of a combination mode at around 4.5\,$\mu$m \citep{1981CP.....56..367H}.
Its contribution is estimated from the column density determined from the absorption at 3.0\,$\mu$m (brown dot-dashed line) and
taken into account in the following fit for Object\,1.  The feature is very broad and does not affect the other features seen in this spectral range except that it adds
additional continuum.  For Object\,2, no information is available from the 3\,$\mu$m region and we assume the same column density of H$_2$O ice
as in Object\,1.  Inclusion of the H$_2$O ice component reduces the column densities of other components by about 10\% for Object\,2 because it changes only the continuum level,
but does not affect the fit and the following discussion.
Deep absorption features are clearly seen at $\sim 4.26$ and $\sim 4.67$\,$\mu$m in both objects and they are attributed to CO$_2$ and CO ices,
respectively.  The intrinsic widths of the two bands are narrower than the IRC spectral resolution ($\sim 0.03$\,$\mu$m) and
the ice properties that are extracted from the details of the band profiles cannot be discussed \citep[e.g.,][]{2003A&A...408..981P}.
In the following fit, we assume that these two bands are approximated by a single Gaussian.  

Around the CO ice absorption, a distinct
shoulder is seen at $\sim 4.6$\,$\mu$m for both objects.  A broad wing is also present at the longer wavelength side.
These can be attributed to a CO gas component, whose $P$ and $R$ branches account for the features 
at the longer and shorter wavelength sides, respectively \citep{2003A&A...408..981P, 2012A&A...538A..57A}.  The XCN feature is also present at around 4.6\,$\mu$m 
and can contribute to the blue shoulder \citep{2005A&A...441..249V}.
We calculate the CO gas at temperatures between 50 to 500\,K with an interval of 50\,K and search for the best fit.
For Object\,1, the blue shoulder is very strong compared to the red wing, and thus the XCN component is required to account for the blue shoulder.  
CO gas is needed to fit the red wing of the 4.67\,$\mu$m feature.
The CO gas temperature is, however, not well constrained for Object\,1 because of the presence of the overlapping XCN feature in the blue side.
CO gas with temperatures between 150 to 500\,K does not make a significant difference in the fit if we adjust the XCN component.
For Object\,2, the red wing component is relatively strong.  The CO gas can account for both the blue shoulder and red wing without XCN component.
Therefore, the XCN component is not included in the fit for Object\,2.
The best fit of the CO gas is obtained with 150\,K.  We assume the same temperature (150\,K) for the CO gas in Object\,1.
The XCN feature consists of two components \citep{2005A&A...441..249V}, but  they are not resolved with the present spectral resolution.
We simply assume a single Gaussian for the XCN component in the fit for Object\,1.

In addition to these components, there is a broad feature at around 4.4\,$\mu$m.  Similar features have been seen towards several YSOs
\citep[e.g.,][]{2010A&A...514A..12S}, but their origin is not known.  \added{The number of of the objects that show the feature is still small,
and any correlation with other features is not recognized.  The feature is not seen in AKARI spectra of
background stars \citep{2013ApJ...775...85N}.}
We call it 4.4\,$\mu$m component and approximate it by a Gaussian. 
Also there are small dips
at around 4.5\,$\mu$m and at a shorter wavelength side 
($\sim 4.1$\,$\mu$m) of the CO$_2$ feature seen in the spectrum of Object\,2.  Their origins are not known at the present either.  We do not include these components in the fit
since they are small features and do not affect the following results.  In the following fits, the peak wavelength and the band width are set as free parameters except for the 4.4\,$\mu$m component
for Object\,2, for which we fix the peak wavelength and width as being the same as those for Object\,1.

The best fit results are shown by the red solid lines in Figure~\ref{fig:CO2}.  The column densities and the peak wavelengths in the best fits are summarized in
Table~\ref{tab:sum}.  For CO$_2$ ice, the peak wavelength is found to be shorter than 4.26\,$\mu$m for both objects (Table~\ref{tab:sum}).  While the 4.4\,$\mu$m broad component shifts the
peak wavelength of the CO$_2$ component slightly, the observed peaks are always seen at a wavelength of 4.26\,$\mu$m or shorter.  
Therefore, the peak is clearly shorter than the peak position of apolar CO$_2$ ice of $> 4.26$\,$\mu$m,
suggesting that a non-negligible fraction of CO$_2$ ice is either polar or pure \citep{1997A&A...328..649E}.  For CO ice, the absorption feature is observed at 4.66--4.67\,$\mu$m, which agrees with pure CO ice.
Polar CO ice has a peak at a wavelength longer than 4.68\,$\mu$m and does not match with the observed spectra \citep{1997A&A...328..649E}. 
The differences in the peak wavelengths are smaller than the spectral resolution ($\sim 0.03$\,$\mu$m), but comparable with the resolution per pixel ($\sim 0.01$\,$\mu$m).
They are larger than the uncertainty in the wavelength ($\sim 0.005$\,$\mu$m, see \S\ref{sec:obs}).  High-spectral resolution
observations are needed to confirm the properties of the CO and CO$_2$ ices accurately.

The column densities of CO and CO$_2$ ices are estimated from the absorption features at 4.67 and 4.26\,$\mu$m, assuming that the band strengths
are 1.1 and $7.6 \times 10^{-17}$\,cm\,molecules$^{-1}$, respectively \citep{1995A&A...296..810G}.  Since both features are deep, and narrower than the spectral resolution,
we need to take account of the saturation effect of the low spectral resolution.  We simulate the features in a way similar to \citet{2010A&A...514A..12S}, assuming that
the band width (FWHM) is 18 and 9.71\,cm$^{-1}$ for the CO$_2$ and CO features, respectively \citep{2004ApJS..151...35G}, 
and estimate the correction factors.   The correction factors are found to be 1.19 and 1.16 for the CO$_2$ absorption feature for Object\,1 and Object\,2, respectively,
and 1.19 and 1.27 for the CO feature for Object\,1 and Object\,2, respectively. Using these correction factors,
we estimate the column densities of CO and CO$_2$ ices as shown in Table~\ref{tab:sum}.

 The XCN feature is known to consist of two components, 4.60 and 4.62\,$\mu$m. The 4.62\,$\mu$m component is securely assigned to OCN$^-$ \citep{2005A&A...441..249V}, 
 while the carrier of the 4.60\,$\mu$m component is unknown.  The present data show a peak at 4.60\,$\mu$m.  However, the peak wavelength depends on the assumed
 CO gas temperature to some extent.  With a higher CO gas temperature, the peak shifts to a longer wavelength.  Since the present spectral resolution is not enough to
 resolve the XCN feature and the CO gas ro-vibrational transitions, it is not possible to discuss the details of the XCN component.
 We assume the same band strengths for both components as $1.3 \times 10^{-16}$\,cm\,molecule$^{-1}$ \citep{2004A&A...415..425V, 2011ApJ...740..109O}
 and estimate the column density.    Since the feature is shallow and broad, we do not apply any correction for the saturation effect.  
 Fits with the CO gas of temperatures different from 150\,K always provide a higher XCN abundance.  Thus, the estimated XCN abundance should be taken as a lower limit.
 The results are summarized in Table~\ref{tab:sum}.
 For the 4.4\,$\mu$m feature, the integrated band intensities  ($=\int \tau$ d$\nu$) in units of cm$^{-1}$ are given in Table~\ref{tab:sum}.

\subsection{5.0--8.0\,$\mu$m} 
\begin{figure}[ht!]
\epsscale{0.6}
\plotone{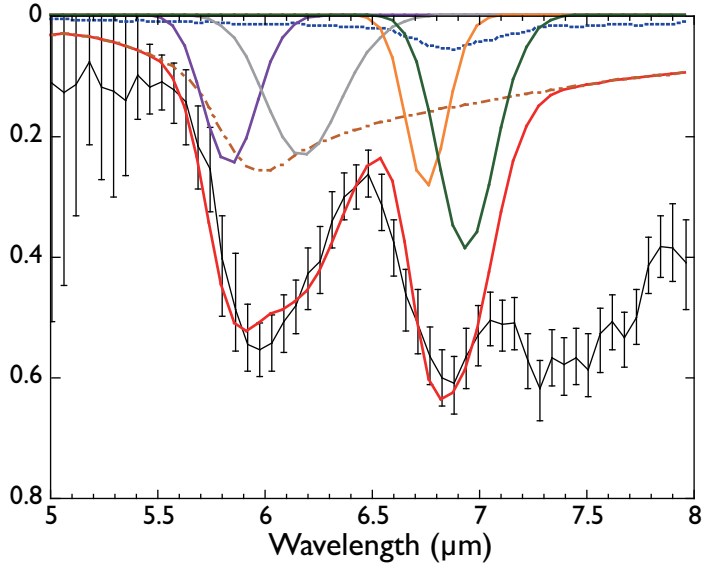}
\caption{ Optical depth spectrum of Object\,1 in 5.0--8.0\,$\mu$m.  The observed spectrum is indicated by the
black solid line with the error bars.    The contribution of H$_2$O ice estimated from the 3.0\,$\mu$m band and that of 
CH$_3$OH ice estimated from the 3.54\,$\mu$m band are indicated by the brown
dot-dashed line and the  blue dotted line, respectively.  
The red solid line shows the best fit result with the summation of the C1, C2, C3, and C4 components for the spectra, from which the contribution from
H$_2$O ice has been removed, but that of CH$_3$OH has not
(see text).  The purple, gray, orange, and green solid lines indicate the C1, C2, C3, and C4 components, respectively.} \label{fig:5_8}
\end{figure}

Figure~\ref{fig:5_8} shows the optical depth spectrum of Object\,1 in the region 5.0--8.0\,$\mu$m, which shows absorption features at around 6 and 6.8\,$\mu$m.
\citet{2008ApJ...678..985B} decompose the ice features in this spectral range into five distinct components:
C1 (5.84\,$\mu$m), C2 (6.18\,$\mu$m), C3 (6.755\,$\mu$m), C4 (6.943\,$\mu$m), and C5 (a 
broad component covering 5.8--8\,$\mu$m).  Possible carriers are discussed in detail in \citet{2001A&A...376..254K}, \citet{2008ApJ...678..985B}, and
\citet{2011ApJ...740..109O}.   The C1 component has been attributed to H$_2$O, H$_2$CO, and HCOOH ices.  We assume the H$_2$O ice
column density from the 3.0\,$\mu$m fit and estimate its contribution (brown dot-dashed line).  The band of H$_2$O ice shifts to a longer wavelength and the peak intensity increases, when
H$_2$O ice is diluted by CO$_2$ \citep{2005ApJ...635L.145K}.  To be consistent with the fit of the 3.0\,$\mu$m band, however,
we use the same pure amorphous H$_2$O ice data at 60\,K in the fit.  There is also a weak feature of CH$_3$OH ice at around 6.8\,$\mu$m.  Its contribution is estimated from the 3.54\,$\mu$m band
(Figure~\ref{fig:Obj1:2-4}), which is found to be insignificant compared to the observed feature (blue dotted line).
After subtracting only the contribution from H$_2$O ice as in  \citet{2008ApJ...678..985B}, we fit the spectrum with the C1, C2, C3, and C4 components 
simply assuming that they are approximated by Gaussians with the central peak wavelengths and the FWHMs given by \citet{2008ApJ...678..985B}  convolved with the IRC spectral resolution 
(0.12\,$\mu$m).   Because of the low spectral resolution of the present spectrum, detailed band profiles do not affect the fit results.
The four-component fit is shown by the red solid line, which reproduces the observed spectrum fairly well.  The present fit does not require the C5 component, but
it is difficult to confirm the presence or absence of the C5 component from the present spectrum because of the uncertainty in the assumed continuum.  
The peak optical depths of the four components in the best fit are shown in Table~\ref{tab:sum}.
The contribution of amorphous H$_2$O ice is estimated to be about a half ($\sim 54\%$) of the 6.0\,$\mu$m feature.
There seems to be also a feature at around 7.3--7.5\,$\mu$m.  It is weak and its presence has to be confirmed by further observations.

\subsection{8.0--13.0\,$\mu$m}

\begin{figure}[ht!]
\plotone{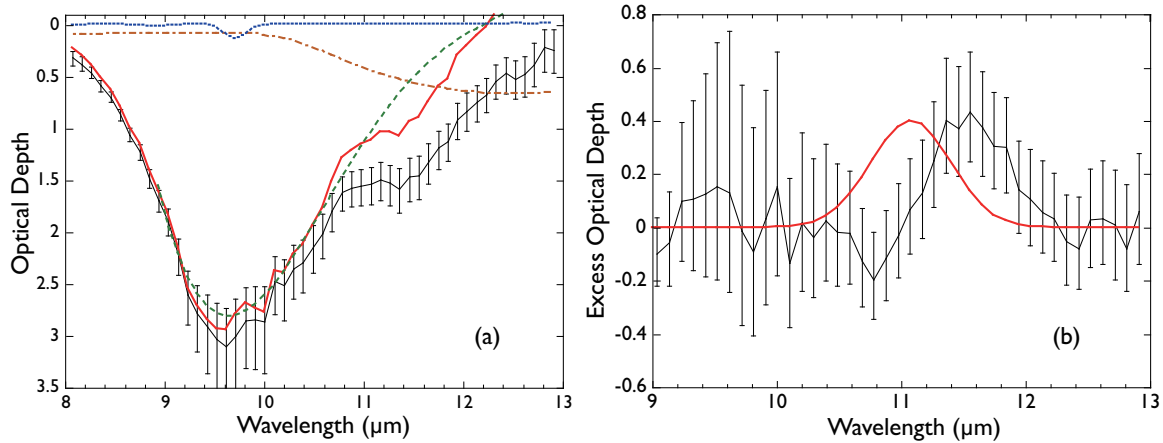}
\caption{(a) Optical depth spectrum of Object\,1 in 8.0--13.0\,$\mu$m.  The observed spectrum is indicated by the
black solid line with the error bars.  The contribution of H$_2$O ice estimated from the 3.0\,$\mu$m band is shown by the brown
dot-dashed line, while that of CH$_3$OH ice from the 3.53\,$\mu$m band is indicated by the blue dotted line.
The red solid line shows the optical depth after subtracting the contributions from H$_2$O ice and CH$_3$OH ice.  The green dashed line shows a polynomial
fit to the red solid line (see text). (b) Excess optical depth after subtracting the polynomial fit to the smooth component (green dashed line in (a)).  
The red line indicates the average excess profile at 11\,$\mu$m derived in  \citet{2020MNRAS.493.4463D}.  Note that the actual profiles have some asymmetry, which
varies from object to object and is not included in the plot (see text).  The amplitude is scaled to that of Object\,1.
\label{fig:10}}
\end{figure}

Figure~\ref{fig:10}a shows the optical depth spectrum of Object\,1 for 8.0--13.0\,$\mu$m together with the contributions from
H$_2$O ice (brown dot-dashed line) and CH$_3$OH ice (blue dotted line) estimated from the 3.0 and 3.53\,$\mu$m bands, 
respectively.  The contribution from CH$_3$OH ice is negligible, while H$_2$O ice adds broad absorption at wavelengths
longer than 10\,$\mu$m.  The spectrum shows deep absorption at around 10\,$\mu$m attributable to amorphous silicate.  We estimate the optical depth at
9.7\,$\mu$m, $\tau_{9.7}$,  as $2.9 \pm 0.3$ after subtracting the contributions from H$_2$O and CH$_3$OH ices.
In addition, there seems some excess absorption at around 11\,$\mu$m on the smooth absorption feature of amorphous silicate.  \citet{2016MNRAS.457.1593W} report
that similar excess is seen towards YSOs and in the interstellar medium (ISM), attributing it to crystalline forsterite.
\citet{2020MNRAS.493.4463D} make a thorough study of this feature in the various lines-of-sight and show the ubiquitous presence of the feature in YSOs and even in the diffuse ISM.  
They discuss several possible origins of the feature
in detail, concluding that it arises from crystalline silicate.
\citet{2020MNRAS.493.4463D} use the spectral regions 9.9--10.2 and 12.0--13.0\,$\mu$m to estimate the amorphous component and extract excess absorption.
Since the spectral region of 9.9--10.2\,$\mu$m of the present spectrum is very noisy, 
we fit instead the spectral regions 8.9--10.3 and 12.0--13.0\,$\mu$m by a quadratic equation and extract the excess absorption (Figure~\ref{fig:10}b).   
The equation fits the spectrum in 9.9--10.2\,$\mu$m reasonably well, and
the different choice of the spectral region for the fit does not affect the extracted excess and the following discussion.
The red line in Figure~\ref{fig:10} indicates the average excess profile derived in  \citet{2020MNRAS.493.4463D}.  Note that the actual profile has some asymmetry,
which varies from object to object.  In Figure~\ref{fig:10}, a Gaussian with the average peak wavelength (11.08\,$\mu$m) and the average FWHM  (0.76\,$\mu$m) 
is simply plotted to show an approximate profile of the average excess.  
The amplitude is scaled to that of Object\,1.
The excess of Object\,1 peaks at around  11.3--11.5\,$\mu$m, which is slightly longer than the peaks derived by \citet{2020MNRAS.493.4463D}, particularly compared
to those found in the ISM ($\sim 11.1$\,$\mu$m).  Note that some YSOs show peaks at longer wavelengths (up to $\sim 11.2$\,$\mu$m).  
The ratio of the excess to the silicate absorption at 9.7\,$\mu$m ($ 0.13 \pm 0.06$) is also larger, but is  roughly in agreement with those found in YSOs
($\sim 0.05$) and the diffuse ISM \citep[$\sim 0.04$,][]{2020MNRAS.493.4463D} within the uncertainty.

\begin{deluxetable*}{cCCCCC}
\tablecaption{Column densities and peak wavelengths of ice species in the best fit \label{tab:sum}}
\tablewidth{0pt}
\tablehead{
\colhead{} & 
 \multicolumn{3}{c}{Object\,1} &  \multicolumn{2}{c}{Object\,2}\\
\colhead{ } & \colhead{Column densities} & \colhead{Abundance\tablenotemark{a} } & \colhead{Peak wavelength} &  \colhead{Column densities} &   \colhead{Peak wavelength} \\
\colhead{Species}  & \colhead{ ($\times 10^{17}$\,cm$^{-2}$)} & (\%) & ($\mu${\mathrm m}) & \colhead{ ($\times 10^{17}$\,cm$^{-2}$)} &  ($\mu${\mathrm m})
}
\startdata 
H$_2$O ice & 54.9 \pm 1.1 & 100 & 3.06 &  \nodata & \nodata\\
CO$_2$ ice &  6.5 \pm 0.2 &  11.8 \pm 0.4 & 4.25 & 5.3 \pm 0.3  & 4.26 \\
CO ice & 12.7 \pm 1.0 & 23.1 \pm 2.0 & 4.66 & 19.5 \pm 1.5  & 4.67 \\
CO gas & 28.0 \pm 5.3 & 51.0 \pm 9.6 & \nodata & 109  \pm 4 & \nodata \\
XCN & 0.9 \pm 0.1 & 1.7 \pm 0.2 & 4.60 & \nodata & \nodata \\
CH$_3$OH ice & 2.6 \pm 0.6 & 4.8 \pm 1.0 & 3.54 & \nodata & \nodata\\
4.4\,$\mu$m feature & 37.8 \pm 1.4\tablenotemark{c} &  \nodata & 4.38 & 63.1\pm 1.8\tablenotemark{c} &  4.38 \\
C1  & 0.25 \pm 0.04\tablenotemark{d}& & 5.84 \\
C2 & 0.23 \pm 0.02\tablenotemark{d} & & 6.18 \\
C3  & 0.29 \pm 0.03\tablenotemark{d} & & 6.755 \\
C4  & 0.39 \pm 0.03\tablenotemark{d} & & 6.943 \\
silicate & 2.9 \pm 0.3 \tablenotemark{d} &  & 9.7 \\
\enddata
\tablenotetext{a}{Relative abundance to H$_2$O ice}
\tablenotetext{b}{Upper limit because of the possible contribution from H$_2$CO ice (see text)}
\tablenotetext{c}{Intended band intensity ($=\int \tau$ d$\nu$) in units of cm$^{-1}$}
\tablenotetext{d}{Peak optical depth}
\end{deluxetable*}

\section{Discussion}\label{sec:discussion}

We found two intriguing objects that show deep ice absorption features in the {\it AKARI}/IRC spectroscopic survey of the Galactic plane.
To investigate the nature of the objects and the location of
the ice species, we discuss their ice properties and infrared spectral energy distributions (SEDs) in the following sections.

\subsection{Properties of absorption features}
Ice absorption features have been observed towards YSOs and used as a good indicator for the identification of YSOs \citep{2005MNRAS.364L..71V, 2008ApJ...686L..99S, 2010A&A...514A..12S, 
2009ApJ...699..150S}.
On the other hand, ice absorption is also observed towards background stars sitting behind quiescent, dense clouds  \citep[e.g.,][]{2005ApJ...635L.145K, 2007ApJ...655..332W, 2011ApJ...729...92B, 
2011ApJ...731....9C,
2013ApJ...775...85N}.  Therefore, they are not secure evidence for the identification of YSO nature of the present objects.   In this section, we
discuss possible evidence for thermal processing on ices, which is not expected for the features in background stars and thus
supports the YSO identification of the objects, based on their spectra.

The 3.0\,$\mu$m H$_2$O ice absorption feature of Object\,1 peaks at 3.06\,$\mu$m, which is longer than amorphous H$_2$O ice at 15\,K
($\sim 3.02$\,$\mu$m, Figure~\ref{fig:Obj1:2-4}a).  The difference is small, but is larger than the uncertainty in the wavelength 
($\sim 0.005$\,$\mu$m).  The absorption profile of 15\,K H$_2$O ice does not fit the absorption profile at the longer wavelength side well, which 
requires a contribution from warm amorphous H$_2$O ice of thermally processed.
The shape effect does not account for the difference and the low spectral resolution does not affect the characteristics of the band profile either.

\begin{figure}[ht!]
\epsscale{0.7}
\plotone{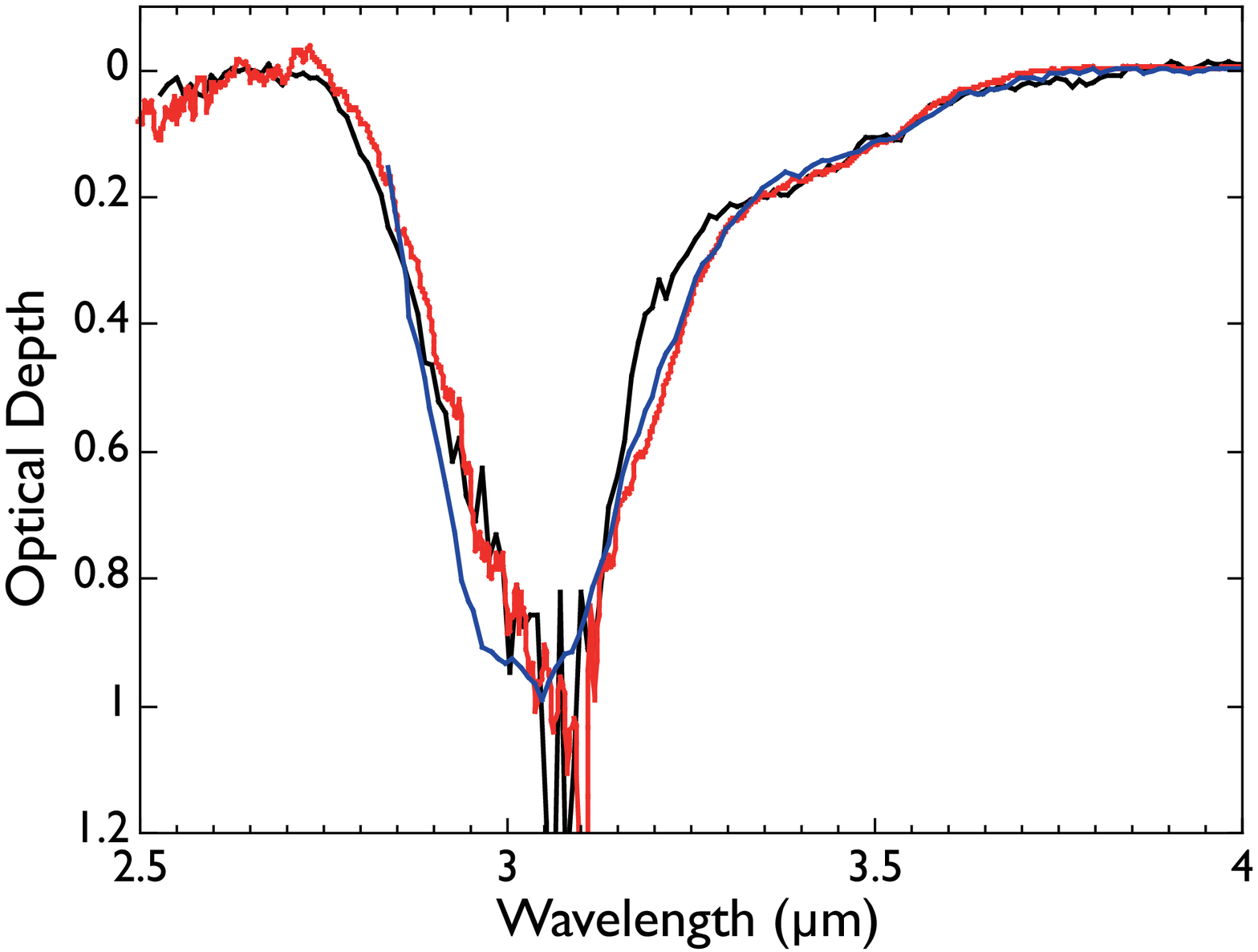}
\caption{Comparison of the 3\,$\mu$m H$_2$O ice optical depth spectrum of Object\,1(black line) with those of a YSO \citep[AFGL\,2136,][]{2004ApJS..151...35G} 
and of a background star \citep[2MASS\,J1872690-0438406,][]{2011ApJ...729...92B}.  The optical depth spectra are normalized to unity at the maximum.
\label{fig:H2O_comp}}
\end{figure}

\citet{2004ApJS..151...35G} analyze the 3.0\,$\mu$m of H$_2$O ice for a number of MYSOs and show that some of them require
a warm ($ \ge 50$\,K) ice component in addition to the 10\,K component.
\citet{2011ApJ...729...92B} presented spectra of the H$_2$O ice absorption at 3\,$\mu$m for several background stars, suggesting no variations in
the band profile.  The 3\,$\mu$m absorption features of background stars have a peak between 3.0 and 3.1\,$\mu$m, similar to Object\,1.
Figure~\ref{fig:H2O_comp} shows a comparison of the 3\,$\mu$m optical depth spectrum of Object\,1 with those of a YSO \citep[AFGL\,2136,][]{2004ApJS..151...35G} and
a background star \citep[2MASS\,J1872690-0438406,][]{2011ApJ...729...92B}.  The background star shows a wider profile in the blue side 
than the YSO and Object\,1.  Other background stars in  \citet{2011ApJ...729...92B} show similar characteristics.  At the red side, no difference is seen between
the YSO and the background star, and Object\,1 shows a slightly narrower width.  The wider width at the blue side of the background star may be attributable to the
presence of cold, amorphous ice.  The H$_2$O ice profile of Object\,1 is in better agreement with the YSO spectrum, but
the difference in the profile between the YSO and the background star is not very large.
It should also be noted that large H$_2$O ice dust shifts the peak to a longer wavelength \citep{1989ApJ...344..413S}.  
While it cannot be ruled out that the longer wavelength of peak absorption of the 3\,$\mu$m feature could be accounted for by
large dust, the 60\,K amorphous ice of a CDE fits the observed spectrum reasonably well, suggesting that 
the observed 3.0\,$\mu$m band profile can also be attributed to thermally processed H$_2$O ice towards Object\,1.

The spectra of 4--5\,$\mu$m contain ample information on the properties of ice species.
The abundance of CO$_2$ ice (relative to H$_2$O ice) is $11.8 \pm 0.4$\% for Object\,1.  It is in the range for
MYSOs, and at the lowest end of the abundance distribution of LYSOs and lower than the range of background stars \citep{2015ARA&A..53..541B}.
Therefore, this abundance ratio suggests that Object\,1 may be a MYSO, although this is not definitive evidence,
taking account of the distribution of the abundance and the uncertainty in the abundance estimation.  

On the other hand, CO ice abundance is large ($23.1 \pm 2.0$\%) for Object\,1 and its column density is large compared to CO$_2$ ice for Object\,2.
This is  a secure result since the observed spectra of both objects show very deep absorption at 4.67\,$\mu$m despite the low spectral resolution.
The CO ice abundance is expected to decrease in a higher temperature environment because of its low sublimation temperature ($\sim 20$\,K).
The median abundance of CO ice in MYSOs is smaller than that in LYSO and background stars,
but the CO ice abundance has a wider distribution than that of CO$_2$ ice \citep{2015ARA&A..53..541B}.
The observed abundance is in a typical range for LYSOs and background stars, while it is still in the range of the abundance distribution of MYSOs
 \citep{2015ARA&A..53..541B}.

The band peak positions of CO ice suggests that it is pure CO ice, while that of CO$_2$ ice suggests that it is either pure or polar.
At an early phase of ice formation, CO$_2$ ice is thought to be formed concurrently with H$_2$O ice and thus it should be polar.  
At some point, most of frozen-out CO is no longer converted into CO$_2$ ice and the abundance of CO ice increases \citep{2011ApJ...740..109O}.
The observed large abundance of CO ice suggests that both objects may be in this evolutionary stage.  
The suggested properties of pure CO and polar CO$_2$ ices are compatible with this interpretation, 
although the present spectra do not have a sufficient spectral resolution to discuss their profiles in detail.  

The presence of warm CO gas component is a strong indicator for the presence of an embedded heating source.  Warm CO gas ($> 50$\,K) has been observed in
absorption towards embedded YSOs \citep{2003A&A...408..981P, 2012A&A...538A..57A}.  In dense clouds, the gas temperature is supposed to be as low
as 20\,K and warm CO gas has not been observed in background stars \citep{2013ApJ...775...85N}.
For Object\,1, the red wing of the 4.67\,$\mu$m band can be best accounted for by warm CO gas ($> 100$\,K).   
For Object\,2, the strong blue
shoulder cannot be attributed to XCN, since it would require a very large column density for the XCN component, suggesting that the presence of warm CO gas ($\sim 150$\,K) is
a secure conclusion.  

CH$_3$OH ice is believed to form on grain surfaces and several different formation processes are proposed by laboratory experiments; i.e., ultraviolet photolysis,
radiolysis, and CO hydrogenation \citep[e.g.][]{1999Icar..140..451H, 2007ApJ...668.1001W}.
Large amounts of CH$_3$OH ice have been observed only towards YSOs and very dense cores that are likely to form stars \citep{2011ApJ...729...92B}, 
and  it could be a good indicator of YSOs \citep{2009ApJ...702L.128A, 2017ApJ...843L..36A}.  
\replaced{However, the CH$_3$OH ice abundance of Object\,1 is not very large and is in the range of YSOs and background stars 
 \citep{2011ApJ...729...92B, 2011ApJ...742...28W}.  Thus, it is not a secure indicator of the identification as a YSO for Object\,1.}
 {However, the CH$_3$OH ice abundance of Object\,1 is in the range of YSOs and such background stars, although a factor of $\sim 5$ less than the 
 most extreme YSO cases.  Thus, it does not seem to be a secure indicator of the identification as a YSO for Object\,1.}
The 4.62\,$\mu$m XCN feature has also been thought to be an indicator of energetic processing.  However, the relatively large abundance observed in
LYSOs suggests that the carrier can be produced through purely thermal acid-base reactions in ices \citep{2004A&A...415..425V, 2005A&A...441..249V}.
It should be noted that the XCN feature has not been observed towards background stars.  The estimated abundance of XCN in Object\,1 is quite large ($1.7 \pm 0.2$\%), being in the range for
MYSOs and much larger than the upper limit for background stars \citep[$<0.5\%$ for OCN$^-$,][]{2011ApJ...740..109O, 2015ARA&A..53..541B}.  The XCN abundance in Object\,1 depends
on the assumed temperature of CO gas, and the present result provides a lower limit (\S\ref{subsec:45}).
Higher spectral resolution data that resolve ro-vibrational bands of the CO gas lines and XCN feature are needed to estimate the
the properties of the  CO gas and the XCN abundance accurately \citep[e.g.,][]{2003A&A...408..981P}.

\begin{figure}[ht!]
\epsscale{0.7}
\plotone{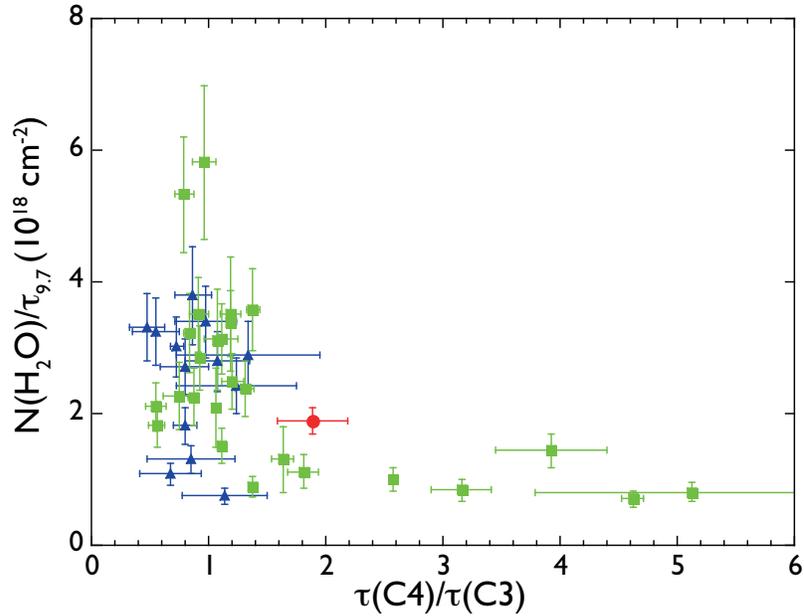}
\caption{ H$_2$O ice column density normalized by the peak optical depth of the silicate absorption at 9.7\,$\mu$m (a measure of H$_2$O abundance)
against the ratio of the integrated optical depths of the C4 to C3 components.  The data point of Object\,1 is shown by the red circle.  The green squares and blue triangles indicate the
data for YSOs and background stars, respectively, taken from \citet{2008ApJ...678..985B, 2011ApJ...729...92B}.
\label{fig:C1}}
\end{figure}

The features in 5.0--8.0\,$\mu$m (C1--C5) can also be used to study the thermal processing of ices \citep[e.g.,][]{2011ApJ...729...92B}.
\replaced{There is no clear difference in the distribution of each of the C1--C4 feature strengths among MYSOs, LYSOs, and background stars,
and the presence of these features can neither support nor rule out the YSO nature of the objects unambiguously \citep{2008ApJ...678..985B, 2011ApJ...740..109O}.}
{The C1--C4 components are present in all classes of objects (MYSOs, LYSOs, and background stars) and their strengths typically correlate well with
the H$_2$O column density.  The mere presence of these feature therefore can neither support nor rule out the YSO nature of the objects unambiguously \citep{2008ApJ...678..985B, 2011ApJ...740..109O}.}
However, the ratio of the C4 to C3 components is large towards \added{some} YSOs, shifting the 6.85\,$\mu$m feature to a longer wavelength, compared to background stars
\citep{2001A&A...376..254K, 2008ApJ...678..985B, 2011ApJ...729...92B, 2009ApJ...690..683R}.  
Figure~\ref{fig:C1} plots the H$_2$O ice column density normalized by the peak optical depth of the silicate band (a measure of the
H$_2$O ice abundance) against the ratio of the integrated optical depths of the C4 to C3 components, $\tau(C4)/\tau(C3)$, for Object\,1 together with those of the YSO and background star samples 
taken from \citet{2008ApJ...678..985B, 2011ApJ...729...92B}.  
Object\,1 shows a relatively low H$_2$O ice abundance and 
the ratio of the C4 to C3 components larger than the majority of background star samples (blue triangles). It is located in the region occupied mostly by
YSOs.
It should be noted that the C5 component has not been observed towards background stars \citep{ 2011ApJ...740..109O}.
The present spectrum is not sufficient to confirm the presence of the C5 component.

The observed silicate absorption at 10\,$\mu$m in Object\,1 shows excess absorption at around 11.3\,$\mu$m and it is best ascribed to crystalline silicate.
The excess is detected in YSOs as well as in the diffuse ISM.  
In the ISM, crystalline silicates are thought to be gradually amorphized by cosmic-ray hits \citep{2007ApJ...662..372B}.  
Therefore, the presence of the excess itself cannot distinguish YSOs from background stars.
A larger amount of excess is sometimes seen towards embedded YSOs, which can be attributed to thermal processing of amorphous silicate by the radiation from the YSO
\citep{2020MNRAS.493.4463D}.  
The observed relatively large excess may suggest MYSO origin.  Further observations of the 10\,$\mu$m band are needed to estimate the amount of the excess accurately.

\begin{figure}[ht!]
\epsscale{0.7}
\plotone{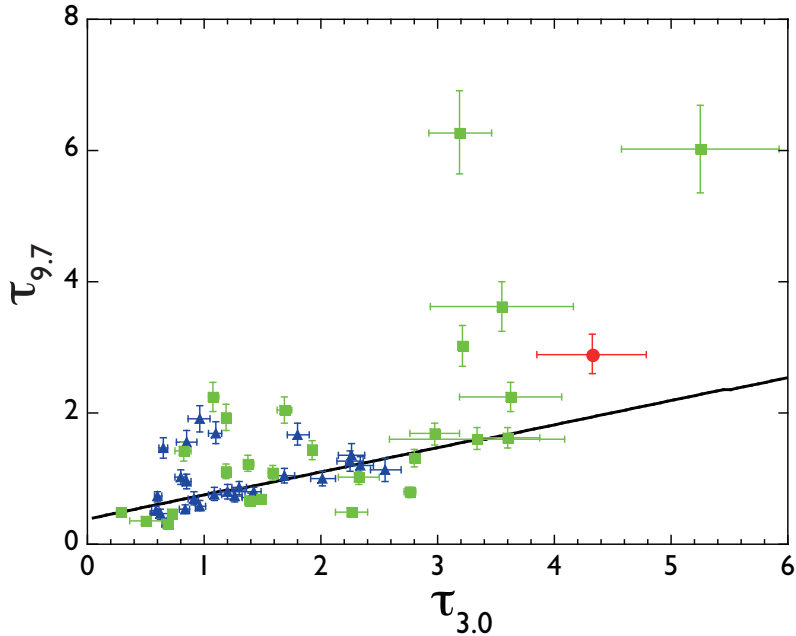}
\caption{ Peak optical depth at 9.7\,$\mu$m $\tau_{9.7}$ against that at 3\,$\mu$m $\tau_{3.0}$.  
The black solid line shows a correlation line for the data of background stars \citep{2011ApJ...729...92B}.
Object\,1 is shown by the red circle.  
The green squares and blue triangles indicate the
data for YSOs and background stars, respectively, taken from \citet{2008ApJ...678..985B, 2011ApJ...729...92B}.
\label{fig:9-3}}
\end{figure}

Figure~\ref{fig:9-3} shows a correlation of the peak optical depth at 9.7\,$\mu$m, $\tau_{9.7}$, against that at 3\,$\mu$m, $\tau_{3.0}$ for the YSO and background star
samples taken also from \citet{2008ApJ...678..985B, 2011ApJ...729...92B}. The correlation line for
the sample of background stars is also indicated by the black solid line  \citep{2011ApJ...729...92B}. 
Most of the background stars lie near the correlation line, while there are several YSOs above the correlation line, in particular for $\tau_{3.0} > 3$.
Object\,1 is located well above the correlation line at a large $\tau_{3.0} $ (= 4.32),  suggesting that Object\,1 may be a YSO.  
Note that the background stars located well above the correlation line at around $\tau_{3.0} \sim 0.6-1.0$ are those towards the core L\,328, which may trace
diffuse ISM rather than dense clouds \citep{2011ApJ...729...92B}.

The difference in abundance of various ice species among MYSOs, LMYSOs, background stars is not very large and their abundance distributions have overlapping
ranges \citep{2011ApJ...740..109O, 2011ApJ...729...92B, 2015ARA&A..53..541B}.  Therefore, it is difficult to draw a definite conclusion on the nature of the two objects
from the abundance of ice species.  On the other hand, there is no evidence against the YSO identification for both objects.   Several lines of evidence indicate that Object\,1 may be a (M)YSO.  
The presence of warm CO gas and XCN feature and the relatively large C4 to C3 component ratio in Object\,1 support the YSO nature together with 
the skewed profile of the 3.0\,$\mu$m H$_2$O ice.
Object\,2 has less evidence because of the limited range of the reliable spectrum, but the strong blue shoulder of the CO ice absorption feature suggests the presence of a large amount of warm
CO gas towards Object\,2, supporting that Object\,2 is also a YSO.   

\subsection{Spectral energy distribution}
The SEDs of both objects peak at around 5\,$\mu$m (Figure~\ref{fig:spectra}).  The SED of embedded YSOs generally increases towards longer wavelengths 
\citep[e.g.,][]{2004ApJS..151...35G, 2008ApJ...678..985B}, while that of background stars peaks at wavelengths shorter than 4\,$\mu$m \citep{2011ApJ...729...92B, 2013ApJ...775...85N}.
There are, however, a few YSOs whose SED has a peak at around 5\,$\mu$m and decreases towards longer wavelengths, resembling the SEDs of Object\,1 and Object\,2
\citep[e.g., B\,35A in ][]{2013ApJ...775...85N}.  

\begin{figure}[ht!]
\epsscale{0.7}
\plotone{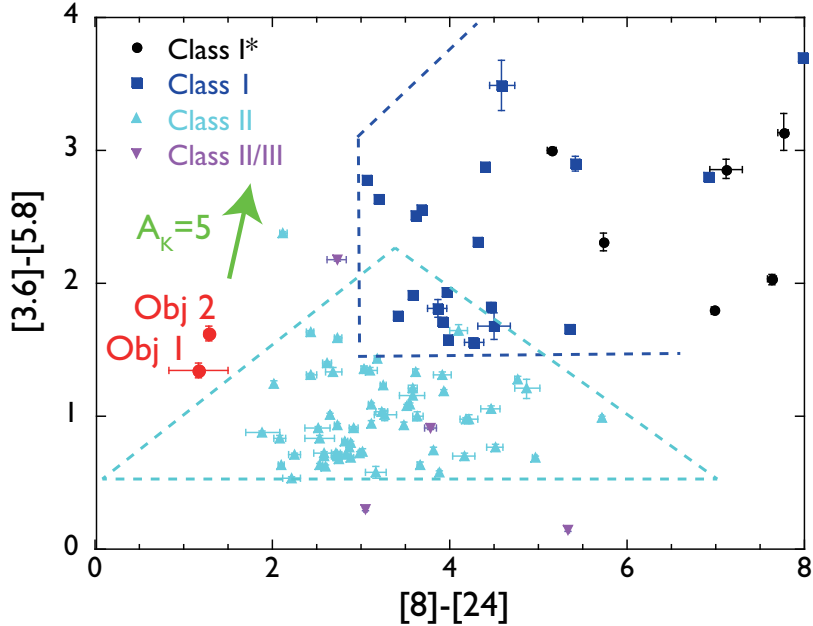}
\caption{{\it Spitzer} IRAC and MIPS [3.6]-[5.8] vs. [8.0]-[24] two-color diagram of YSOs in the NGC\,1333 region \citep{2008ApJ...674..336G}.  
The different color symbols indicate different evolutionary stages of YSOs (black: circles: Class I*, blue squares: Class I, light blue triangles: Class II; purple 
reverse triangles: Class II/III).  
The areas of  YSOs in the Cygnus X region \citep{2010ApJ...720..679B} are also indicated by the dashed lines.  The thick and light blue dashed lines roughly enclose the areas of
Class I  YSOs and Class II YSOs, respectively, in the Cygnus X region.  The positions of Object\,1 and Object\,2 are indicated by the red circles labeled
as Obj\,1 and Obj\,2, respectively.
The green arrow indicates the reddening vector of $A_\mathrm{K}$=5.
\label{fig:2color}}
\end{figure}

Figure~\ref{fig:2color} shows a two-color diagram of {\it Spitzer} IRAC and MIPS [3.6]-[5.8] versus [8.0]-[24] for YSOs in the NGC\,1333 region taken from \citet{2008ApJ...674..336G}.
The distribution of YSOs in the Cygnus X region  \citep{2010ApJ...720..679B} is also indicated by the dashed lines.  Object\,1 and Object\,2 are located outside of
the YSO regions in the two-color diagram. 
They are too blue in 8--24\,$\mu$m compared to standard YSOs.  The reddening vector shown by the green arrow suggests that they are rather background 
stars with large extinction.   Taking the relation $A_\mathrm{v}/\tau_{9.7} = 18.5 \pm 2.0$ \citep{2003ARA&A..41..241D},
the visual extinction $A_\mathrm{v}$ is estimated as $54 \pm 8$.  

The {\it WISE} colors also confirm the background characteristics of both objects.  The {\it WISE} colors of Object\,1, $W1 - W2 = 1.927 \pm 0.030$ and $ W3 - W4 = 0.927 \pm 0.074$, are located well outside of
the YSO region in the classification scheme of  \citet{2014ApJ...791..131K}.  The color $W1 - W2$ is very red, but $W3-W4$ is not, suggesting that it is a background star with large extinction.  
Object\,2 has $W1 - W2 = 1.697 \pm 0.031$ and
$ W3 - W4 $ ($=2.601 \pm 0.057$), which place it in the YSO region.  However, as described in \S\ref{sec:results}, $W4$ is much brighter than MIPS\,24\,$\mu$m data (Figure~\ref{fig:spectra}).
It may have to be taken as an upper limit because of the presence of nebulosity.  If we use MIPS\,24\,$\mu$m data as a replacement of $W4$, then Object\,2 will be
placed outside of the YSO region.

The visual extinction towards the regions of both objects is, however, estimated to be less than 10 based on 
the optical data \citep{2005PASJ...57S...1D}
and no thick CO clouds have been detected towards them \citep{1987ApJ...322..706D, 1989ApJS...71..481B}.  No dark clouds are listed at the positions of the two objects in the {\it Spitzer} dark cloud
catalog either \citep{2009A&A...505..405P}.  Faint nebulosity is present around Object\,2 (\S\ref{sec:results}), whose surface brightness is estimated as 10\,MJy\,sr$^{-1}$ at 24\,$\mu$m.
According to the ISM dust emission model of \citet{2011A&A...525A.103C}, this brightness 
corresponds to the hydrogen column density of $\sim 5 \times 10^{22}$\,cm$^{-2}$ or $A_\mathrm{V}$ of about 25, if the incident radiation field intensity $U$ is similar to the solar neighborhood.  If the radiation field is stronger, $A_\mathrm{V}$ becomes smaller.  The geometry of the nebulosity relative to Object\,2 is not known and 
it is not clear if ice species can be formed in the nebulosity.
No nebulosity is seen around Object\,1 (Figure~\ref{fig:location}).  While the nebulosity could be the origin of the extinction for Object\,2 in part, it seems unlikely that Object\,1 is a background star with large extinction
unless there is an unknown very small, but thick cloud on the line-of-sight.

\begin{figure}[ht!]
\epsscale{0.7}
\plotone{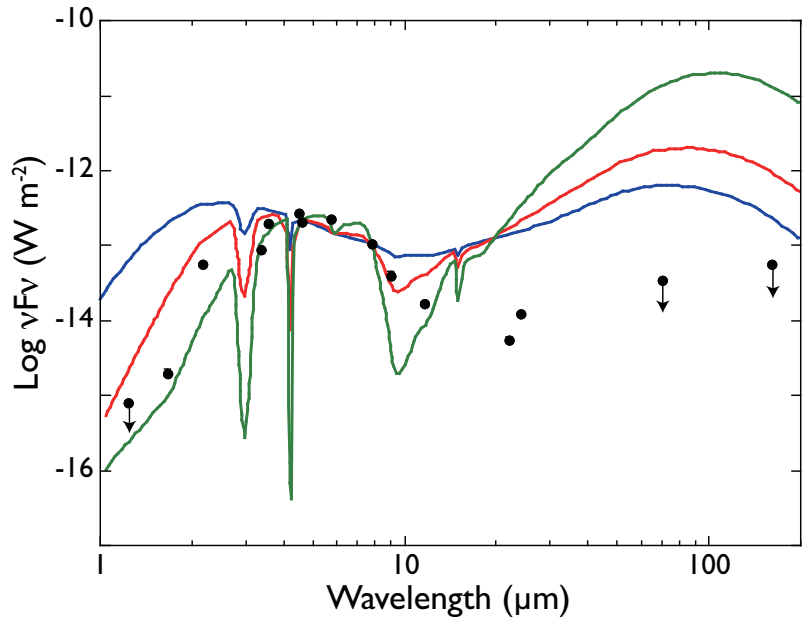}
\caption{SED of Object\,1 and examples of edge-on disk models \citep{2008A&A...486..245C} in units of $\nu F_\nu$.  The black circles show the observed photometric points.
Note that the data at 1.2, 70, and 160\,$\mu$m are upper limits.  The edge-on models of the inclination angle of $45\degr$ are scaled at 4.6\,$\mu$m and shown by the solid lines
for the envelope masses of 0.55 (blue), 1.5 (red), and 4 M$_\odot$ (green).
\label{fig:SED}}
\end{figure}

The blue nature of the objects is also confirmed by the comparison with YSO SED models by \citet{2017A&A...600A..11R}.  No YSO models fit the observations
of both objects satisfactorily.  The best results are obtained with background stars with large extinction ($A_\mathrm{v} \sim 50$) .  Since the models by  \citet{2017A&A...600A..11R}
do not include ice absorption, we compare the present observations with edge-on disk models, in which ice absorption is included \citep{2008A&A...486..245C},
to understand the origin of the discrepancy.  Detailed modeling is not the purpose of this paper and
only comparison of the SED of Object\,1 with some example models is shown in Figure~\ref{fig:SED}.  
The same discussion can be applied for Object\,2.

In Figure~\ref{fig:SED}, the edge-on disk models of the inclination angle of $45\degr$ with different envelope masses \citep{2008A&A...486..245C} are shown by the 
solid lines together with the observations (black circles).    Models with other inclination angles show a similar trend. 
The blue, red, and green solid lines show examples of the models with the envelope masses of  0.55, 1.5 and 4\,M$_\odot$, respectively, normalized at
4.6\,$\mu$m.
The observed depth of ice absorption can be best reproduced by the model with the envelope mass of 1.5\,M$_\odot$
(red line) among them and it fits the observations up to 12\,$\mu$m generally well.  However, for wavelengths longer than 20\,$\mu$m, the model SED starts deviating from the observations.
The  model SED is either flat (blue) or increasing (red and green),
while the observed SED decreases.  Note that both objects are not detected in the HIGAL survey \citep{2016A&A...591A.149M} and the plotted upper limits are very conservative.  
If we decrease the envelope mass (blue line), it slightly reduces the discrepancy, but the ice absorption features become shallower and does not match
the observations very well.  Models with smaller inclination angles show shallower ice absorption and do not reproduce the observations.  

The edge-on disk with a large inclination angle is optically thick at wavelengths shorter than 5\,$\mu$m, but becomes optically thin at longer wavelengths.  The absorbed energy in the disk
is emitted at wavelengths longer than 20\,$\mu$m.  Therefore, it is a natural consequence that edge-on disk models with large inclination angles and deep absorption show strong
MIR and FIR emission relative to the NIR.  Absorption occurs on the line-of-sight, while emission comes from the entire envelope.  If the absorbing envelope is very clumpy and
located just on the line-of-sight, the MIR to FIR emission is reduced, which could be reconciled with the observed SEDs.   Asymmetric, clumpy distributions of dust are sometimes
seen in proto-planetary disks and transition disks \citep[e.g.,][]{2006ApJ...644L.133F, 2013Sci...340.1199V} and thus may not be unusual for YSOs. 
If the outer radius is truncated, it could also reduce the FIR emission.  Further investigations on modeling are needed to understand the SEDs of
the present objects.

\subsection{Nature of the objects and implications}
Both objects show features at both the blue and red sides of the CO ice feature at 4.67\,$\mu$m and they are attributed to the absorption of warm CO gas.  Warm CO gas is not expected to be present in
quiescent, dense clouds and thus is strong evidence for the YSO identification \citep{2003A&A...408..981P, 2012A&A...538A..57A, 2013ApJ...775...85N, 2016ApJ...829..106O}.
The relatively large abundance of XCN estimated for Object\,1 also supports the YSO characteristics since XCN has never been observed in background stars and the estimated abundance is rather high
compared to the upper limits for background stars \citep{2011ApJ...740..109O, 2015ARA&A..53..541B}.  Further observations with high spectral resolution are needed 
to resolve the CO gas and XCN features and estimate their properties unambiguously.  There are other pieces of evidence to support the YSO
identification for Object\,1, including an indication of thermally processed amorphous H$_2$O ice and the abundance of CO$_2$ ice, but none of them
is very secure evidence since similar characteristics are also observed in background stars.  There is no strong evidence against the
YSO identifications for the both objects in the characteristics of ice absorption features.

On the other hand, the SEDs of both objects are rather blue in 8--24\,$\mu$m, which put them outside of the standard YSO region in the
two-color diagram (Figure~\ref{fig:2color}).  The blue nature and non-detection at FIR suggest that they are background stars with large extinction.
Since there is no evidence for the presence of dense clouds in optical and CO observations, those clouds
must be very compact.  The presence of such isolated dense, compact clouds is not known.  If the two objects are background stars, it will make an impact
on our view of the dense gas distribution in our Galaxy and low-temperature chemistry in the ISM.

H$_2$O ice has also been observed in O-rich AGB stars with high mass-loss rates (OH/IR stars) and post-AGB stars with dense, cooled envelopes 
\citep[e.g.,][]{1976ApJ...207..780G, 1990ApJ...355L..27O, 1999A&A...352..587S}.
Their SEDs have a peak at around 10\,$\mu$m and decreasing towards longer wavelengths \citep{1990ApJ...355L..27O, 2000A&A...364..170D}.
The WISE color $W2 -W3$ of Object\,1 and Object\,2 is $0.129 \pm 0.039$ and $0.687 \pm 0.044$, respectively, which is very blue compared to the $W1-W2$ color, and only
a few evolved stars are located in this region of the two-color diagram \citep{2014ApJ...791..131K}.  The 3\,$\mu$m H$_2$O ice features in those O-rich evolved stars show a narrower profile than those of YSOs
\citep{1976ApJ...207..780G} or have peaks at 3.1\,$\mu$m, indicating that a significant fraction of H$_2$O ice is crystalline \citep{2003MNRAS.345..912M}. While recent models show that 
complex organic ices can be formed in C-rich evolved stars \citep{2021MNRAS.501..491V},
features of other ice species have not been detected in O-rich evolved stars with H$_2$O ice absorption probably because of carbon-poor environments \citep{1999A&A...352..587S}.
Therefore, it is not very likely that the present objects are evolved stars.   Absorption features of various ice species are also detected in starburst galaxies and dust-enshrouded AGNs
\citep[e.g.,][]{2000A&A...357..898S, 2003A&A...402..499S, 2004ApJS..154..184S, 2010ApJ...721.1233I, 2013ApJ...773L..37Y, 2015ApJ...807...29Y}.  They are usually associated with 
the 3.3\,$\mu$m emission of aromatic species and/or emission of hydrogen recombination lines.  Since the present objects are located on the Galactic plane and do not show
these emission features,  it is not likely that they are galaxies either.

If the background star is M-type, it could account for the observed absorption of CO gas.  However, M-type giants show a much broader CO absorption feature
starting from 4.3\,$\mu$m 
and extending up to 4.9\,$\mu$m \citep{2002A&A...394..539H} due to the high temperature of the CO gas in their atmosphere ($\gtrsim 3000$\,K).   M-type dwarfs show the dominance of higher-level transitions in
the first overtone of CO absorption in the 2\,$\mu$m region \citep{2005ApJ...623.1115C} and are expected to have a broad
fundamental transition feature of CO absorption from 4.3 up to $\sim 5$\,$\mu$m \citep{2002A&A...396..967P}.  
Brown dwarfs also show a very broad CO absorption at around 4.5\,$\mu$m \citep{2010ApJ...722..682Y}.  In the spectral region of 7--8\,$\mu$m, M-type giants have a broad
absorption feature of the fundamental transitions of SiO gas \citep{2002A&A...394..539H}, while M-type dwarfs show no strong features \citep{2006ApJ...648..614C}.
SiO gas absorption could be responsible for part of the observed features at 7--8\,$\mu$m in Object\,1 (Figure~\ref{fig:5_8}), but SiO gas has much stronger, broad absorption of the first overtone
at around 4.2\,$\mu$m in the spectra of M-type giants \citep{2002A&A...394..539H}, which is not observed in the present spectrum. 
These expected characteristics do not match with the observed spectra of both objects, suggesting
it unlikely that they are background M-type stars.

Both objects are not located in known star-forming regions.
Figure~\ref{fig:location} indicates that the nearby star-forming activities are located in the southern part of the sky of the two objects, where intense diffuse MIR emission is observed.
According to the catalog of the star-forming regions of \citet{2002ARep...46..193A}, the nearest star-forming region is IRAS\,14004-6104 for Object\,1 (6$\farcm$8 separation)
and  IRAS\,14004-6104 and RAGL\,4188 for Object\,2 (both have a 4$\farcm$2 separation).  The nearest \ion{H}{2} region is G311.540+00.319 for both objects and the separations are
5$\farcm$99 and 4$\farcm$22 for Object\,1 and Object\,2, respectively \citep{2017A&A...597A.114B}.  Such separations may not be unexpected from the drift motion of YSOs
from star-forming regions \citep{1996ApJ...468..306F, 2014MNRAS.439.1829C} and the two objects could be run-away YSOs, 
although the distance to the two objects and their evolutionary stages are not known.  
An isolated YSO is also reported to be present, which may have been formed {\it in situ} \citep{2016MNRAS.460.2822B}.
YSO population not associated with star-forming regions has an impact on the understanding of star-forming activities and history in our Galaxy
\citep[e.g.][]{2005A&A...437..247D, 2019A&A...624A..66R}. 
If they are YSOs, the present result suggests that similar objects have eluded past photometric surveys of YSOs \citep[e.g.,][]{2008ApJ...674..336G, 2009ApJS..184..172G, 2010ApJ...720..679B,
2012AJ....144..179K, 2014ApJ...791..131K}.  
This makes significant implications on the population of YSOs in our Galaxy and nearby galaxies.  
If Object\,1 is a MYSO as suggested from the low CO$_2$ ice abundance, it could further make implications on the formation scenario of massive stars, which is not well understood yet,
and on the galaxy evolution
\cite[e.g.,][]{2007ARA&A..45..481Z, 2019A&A...624A..66R}.
Their SEDs may be  accounted for by clumpy distribution of absorbing materials around a YSO.  Further investigations on modeling are strongly encouraged to understand their nature.  

\section{Summary}\label{sec:summary}
We have discovered two intriguing infrared objects in the {\it AKARI}/IRC slit-less spectroscopic survey of the Galactic Plane.  For Object\,1, a full spectrum from
2.5 to 13\,$\mu$m was successfully extracted, while only a spectrum of 3.1--5\,$\mu$m was reliably extracted for Object\,2 due to the presence of nearby objets
and faint nebulosity.
Both objects show deep absorption features
of H$_2$O, CO$_2$, and CO ices.  They also show warm ($>100$\,K) CO gas absorption, suggesting that they are embedded YSOs.  The spectrum of Object\,1
also indicates strong XCN absorption, further supporting the YSO identification.  Other indicators for the thermal processing of ices are also suggested, though they
are also compatible with the ice properties seen in background stars.  They are not definite evidence for the YSO identification, while there are no indications against it
in their spectra.

On the other hand, their SEDs peak at around 5\,$\mu$m and decrease towards longer wavelengths.  They are not detected in the HIGAL survey.  These characteristics
are better reproduced by background stars with large ($A_\mathrm{V} \sim 50$) extinction.  Both objects are located neither in known star-forming regions nor in
known dense clouds.  Although the observed SEDs may be explained if the absorbing ice species are located in clumpy concentrations just on the line-of-sight in the
edge-on disk surrounding a YSO, their true nature remains uncertain based on the currently available data.

If they are background stars with large extinction, there must be unknown compact, dense clouds in the lines-of-sight towards them.  The presence and the formation of
such isolated dense clouds are not known.  If ices are formed in these environments, it will have an impact on the chemical processes in the ISM.
If they are truly YSOs, their blue color in the MIR suggests that similar kinds of objects have eluded in past photometric surveys, which have significant 
implications on our understanding of star-formation process and the distribution of star-forming activities in our Galaxy and nearby galaxies. It is important to identify true nature of the objects
by further observations.

\acknowledgments
This work is based on observations with {\it AKARI}, a JAXA project
with the participation of ESA.  The authors thank all the members of the {\it AKARI} project
for their continuous support.  They also thank A. C. Adwin Boogert for providing us with the data of the YSO and background star samples and for useful comments, and Issei Yamamura for helpful
comments on the infrared spectrum of M-type stars.
This work is based in part on observations made with the {\it Spitzer Space Telescope}, which was operated by the Jet Propulsion Laboratory, California Institute of Technology under a contract with NASA.
This publication makes use of data products from the Two Micron All Sky Survey, which is a joint project of the University of Massachusetts and the Infrared Processing and Analysis Center/California Institute of Technology, funded by the National Aeronautics and Space Administration and the National Science Foundation, and those from the {\it Wide field Infrared Survey Explorer}, which is a joint
project of the Unviersity of California, Los Angeles, and the Jet Propulsion Laboratory/California Institute of Technology, funded by the National Aeronautics and Space Administration.
{\it Herschel} is an ESA space observatory with science instruments provided by European-led Principal Investigator consortia and with important participation from NASA.
This work has also made use of the SIMBAD database, operated at CDS, Strasbourg, France.
The authors thank the Unviersity of Tokyo Research Internship Program (UTRIP) student Ingrid Koch for her contribution at the initial stage of this work.
 This work was supported by JSPS KAKENHI Grant Number JP16H00934, JP18K03691, JP19H05067, and JP20H05845.
%

\vspace{5mm}
\facilities{AKARI (IRC), Spitzer (IRAC and MIPS), WISE, Herschel (PACS and SPIRE)}






\bibliography{YSO-k}{}
\bibliographystyle{aasjournal}



\end{document}